\documentclass[
a4paper,
10pt,
aps,prl,
onecolumn,
amsmath,amssymb,
superscriptaddress,
nobibnotes,
]{revtex4-2}
\usepackage[utf8]{inputenc}
\usepackage[T1]{fontenc}
\usepackage{graphicx}
\graphicspath{ {./figures/} }

\usepackage{tabularx}
\usepackage{varwidth}

\usepackage{amsthm}
\usepackage{mathdots}
\usepackage{float}
\usepackage{eurosym}

\usepackage[normalem]{ulem}

\usepackage{xcolor}
\usepackage{comment}
\usepackage{siunitx}
\usepackage{makecell}

\DeclareMathOperator*{\argmax}{arg\,max}
\renewcommand\vec{\mathbf}

\usepackage{tikz}
\usetikzlibrary{arrows.meta}

\begin{document}

\title{Higher-order interactions reveal synergistic backbones\\of cycling infrastructure networks}

\author{Christoph Steinacker}
\email[Corresponding author: ]{christoph.steinacker@tu-dresden.de}
\affiliation{Chair for Network Dynamics, Center for Advancing Electronics Dresden (cfaed) and Institute of Theoretical Physics, TUD Dresden University of Technology, Germany}

\author{Henrik Wolf}
\affiliation{Chair for Network Dynamics, Center for Advancing Electronics Dresden (cfaed) and Institute of Theoretical Physics, TUD Dresden University of Technology, Germany}
\affiliation{AMOLF, Amsterdam, Netherlands}

\author{Marc Timme}
\affiliation{Chair for Network Dynamics, Center for Advancing Electronics Dresden (cfaed) and Institute of Theoretical Physics, TUD Dresden University of Technology, Germany}
\affiliation{Center Synergy of Systems (SynoSys), TUD Dresden University of Technology, Germany}

\author{Malte Schr\"oder}
\email[Corresponding author: ]{malte.schroeder@tu-dresden.de}
\affiliation{Chair for Network Dynamics, Center for Advancing Electronics Dresden (cfaed) and Institute of Theoretical Physics, TUD Dresden University of Technology, Germany}

\begin{abstract} \noindent
Infrastructure networks essentially underlie human mobility and transport.
Improving the quality of single links increases network performance locally.
However, efficient transport requires high-quality connected corridors across multi-link paths that do not emerge from independent single-link upgrades.
Here, we introduce a framework for evaluating the impact of jointly upgrading multiple links as inherently higher-order interactions, enabling us to quantify link synergies in complex transport networks.
Two links are synergistic if an upgrade of one increases the benefit of upgrading the other, promoting upgrades of topologically complementary links along the same path while discouraging upgrades of redundant parallel links.
By expressing these synergies as second-order derivatives of overall network performance, we develop an efficient computational framework to identify synergistic links that form a connected network backbone.
We apply our theoretical framework by combining empirical street network and cycling demand data for Hamburg, Germany, with a perturbed utility route choice model for urban bicycle traffic.
Our results reveal synergies from higher-order interactions, thereby enabling strategic infrastructure planning that goes beyond local link importance in complex transport and flow networks.
\end{abstract}

\maketitle

\section*{Introduction}
Human mobility relies on infrastructure networks to enable fast and efficient transport across all scales \cite{Barthelemy2011_SpatialNetworks, Barbosa2018_HumanmobilityModels, Gonzalez2008_UnderstandingHumanPatterns}. Rail networks are essential for long-distance and urban public transport \cite{Sen2003_IndianRailSmallworld, Latora2002_BostonSubwaySmallworld}, street networks enable car traffic in and between cities \cite{Gastner2006_OptimalSpatialDistrNetworks, Colak2016_UnderstandingCongestedTravel, Barthelemy2008_ModelingUrbanStreet}, and dense networks of foot- and bike paths promote short-distance active mobility \cite{Helbing1997_EvolutionHumanTrail, Rhoads2023_SidewalknetworksReview}. While optimal structures emerge from the intrinsic dynamics in some of these networks \cite{Helbing1997_EvolutionHumanTrail, Verma2016_EmergenceCorePeripheries, Mittal2024_InformalPublicTransport}, most networks are explicitly designed to enable direct trips with minimal investments \cite{Gastner2006_OptimalSpatialDistrNetworks, Barthelemy2011_SpatialNetworks}. However, designing efficient infrastructure networks constitutes a complex problem that is often framed as a leader-follower or Stackelberg game \cite{CepedaValero2025_StackelbergInfrastructurePlanning, Yang1998_ModelsAlgorithmsDesign}: transport planners implement network upgrades, and users choose their routes in reaction to these upgrades. Planning network expansions based only on the importance of single links and their current usage, therefore, potentially reinforces inefficient network structures and suboptimal traffic patterns \cite{Wassmer2026_SocialCostGradient, Steinacker2022_DemandDrivenDesign}. Efficient design depends on the interactions and the mutual feedback between the structure of the network and its usage. New infrastructure improves travel times for users, affecting their route choices and the overall amount of traffic within the network; different route choices and induced travel demand in turn impact travel times and overall network performance \cite{Braess1968_BraessParadoxon, Fosgerau2023_BikeabilityInducedDemand}. Even if improving a single link appears beneficial in isolation, it may thus indirectly influence the importance of other links and alter network performance with potentially unintended consequences, as famously illustrated by the classic Braess paradox \cite{Braess1968_BraessParadoxon, Youn2008_PriceanArchyTransportation, Witthaut2012_BraesssOscillatorNetworks,
Motter2018_AntagonisticPhenomenanNetwork, Case2019_BraessMicrofluidNetworks, Schaefer2022_UnderstandingBraessParadox}.

\begin{figure}[!h]
    \centering
    \includegraphics{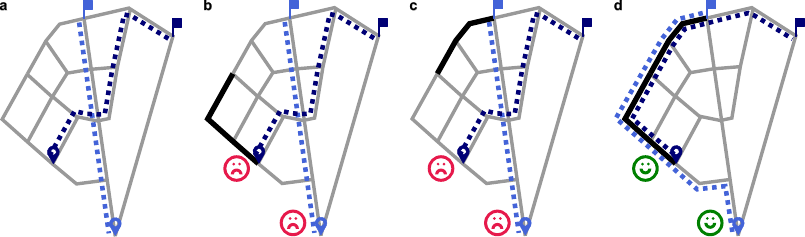}
    \caption{
    \textbf{Higher-order interactions impact the effectiveness of network upgrades.}
    Schematic illustration of the importance of higher-order interactions between network structure and usage.
    (a) In a network without any dedicated infrastructure (gray), users choose the fastest routes (thick dashed lines) from their origin (pins) to their destination (flags).
    (b,c) Upgrading individual segments (thick black lines) is not sufficient to induce route choice changes and may thus not improve the overall network performance at all. The individual importance of both segments is low.
    (d) In contrast, upgrading both segments together changes the route choice decisions and significantly improves network performance for both users simultaneously.
    }
    \label{fig:FIG1_SchematicNetworkImprovement}
\end{figure}

The feedback between network structure and usage naturally gives rise to higher-order interactions between different infrastructure elements (Fig.~\ref{fig:FIG1_SchematicNetworkImprovement}). Higher-order interactions beyond individual links have already proven essential for a wide range of dynamical processes in complex networks \cite{Lambiotte2019_NetworksoHigherorderModels, Battiston2020_NetworksPairwiseInteractions, Battiston2021_PhysicsHigherorderInteractions, Majhi2022_DynamicsHigherorderNetworks, Bick2023_HigherOrderNetworks}, from understanding network connectivity and percolation \cite{Sun2021_HigherorderPercolationHypergraphs, Sun2023_DynamicPercolationNetworks, Bianconi2024_TheoryPercolationHypergraphs, Nortier2025_HigherorderShortestPaths} to revealing causal structures in temporal networks \cite{Scholtes2014_CausalitydrivenTemporalNetworks, Scholtes2016_HigherorderAggregateNetworks, Lambiotte2019_NetworksoHigherorderModels} and explaining social group dynamics \cite{AlvarezRodriguez2021_EvolutionaryDynamicsHigherorder, Majhi2022_DynamicsHigherorderNetworks, Iacopini2024_DynamicsGroupNetworks, Battiston2025_HigherOrderInteractions}. Transport processes across networks inherently involve multiple links along different paths, making this perspective of higher-order interactions especially relevant in the context of mobility infrastructure \cite{Xu2016_HigherorderDependenciesNetworks, Lambiotte2019_NetworksoHigherorderModels, Kim2024_ShortestPathPercolation, Kim2025_ModelingResourceConsumption}. However, many planning approaches still rely solely on first-order individual link importance measures to prioritize network upgrades \cite{Colak2016_UnderstandingCongestedTravel, Kirkley2018_BetweennessCentralityNetworks, Olmos2020_DataScienceFramework}. While dynamical evaluation of individual link importance during network expansions accounts for the impact of already constructed infrastructure \cite{Steinacker2022_DemandDrivenDesign, Paulsen2024_WelfareOptimalExpansionInducedDemand}, explicitly quantifying interactions among two or more links to identify mutually beneficial network upgrades remains an open problem.

Here, we quantify synergies emerging from the simultaneous upgrades of multiple links in an infrastructure network by explicitly accounting for these higher-order interactions. Whereas we identify the importance of a single link with the first-order derivative of the overall network performance with respect to the link's quality, synergies among two or more links are directly captured by second- and higher-order derivatives. These synergies represent the nonlinear and nonlocal interactions among individual links, network usage, and overall performance. We develop a general mathematical framework to evaluate these synergies and identify sets of links that benefit from joint upgrades. In contrast to first-order optimization of network performance, network expansion strategies maximizing these synergies intrinsically form connected network backbones in a self-organized way. Explicitly characterizing higher-order interactions thus offers a broader perspective on quantitatively evaluating flow and transport networks beyond individual link importance and may help prioritize infrastructure upgrades by identifying efficient combinations of network expansions.

\subsection*{Bike network planning}
\begin{figure}[!h]
    \centering
    \includegraphics[width=\linewidth]{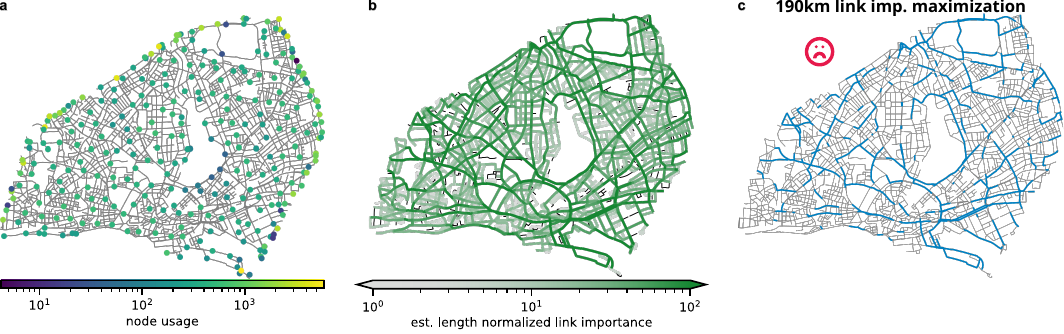}
    \caption{
    \textbf{Network upgrades based on individual link importance remain disconnected.}
    (a) Model travel demand on the street network of Hamburg, Germany, with $287$ origin-destination points (see Methods and Supplementary Notes 1 and 2 for details on the network and demand setting). The color of each point denotes the total number of incoming and outgoing trips. Points on the boundary represent trips beginning and/or ending outside the simulation area.
    (b) First-order link importance representing the estimated length-normalized change of the network performance due to the improvement of each individual link [see Eq.~\eqref{eq:first_derivative_linkimportance} below and Methods for details]. Upgrading a link always improves network performance, with links with higher flows and larger potential quality improvements providing a larger benefit (darker green).
    (c) However, planning network upgrades based on this individual link importance results in a disconnected set of upgrades ($\SI{190}{\km}$ bidirectional links with the highest individual importance, blue).
    }
    \label{fig:FIG2_HamburgFirstOrder}
\end{figure}

\vspace{-3mm}
We illustrate our approach by applying it to bicycle travel on the street network of Hamburg, Germany (Fig.~\ref{fig:FIG2_HamburgFirstOrder}), employing a recently introduced perturbed utility route choice model \cite{Fosgerau2022_PerturbedUtilityRoute, Fosgerau2023_BikeabilityInducedDemand, Fosgerau2026_Sensitivity} to describe network usage. A full explanation of the data, demand setting, and the perturbed utility route choice model is provided in the Methods with detailed parameters and additional visualizations in Supplementary Notes 1-3.

Transport infrastructure design is becoming increasingly important in the context of efficient cycling infrastructure to support the ongoing shift towards more sustainable and active urban mobility \cite{Creutzig2016_UrbanInfrastructureClimate, Olmos2020_DataScienceFramework}. Various planning approaches have been suggested recently, ranging from purely structural models that employ optimal percolation aimed at increasing network connectivity \cite{NateraOrozco2020_DatadrivenStrategiesOptimal, Sebastiao2026_TradeDirectnessCoverage}, to data-driven approaches growing optimal cycling infrastructure based on current network usage \cite{Olmos2020_DataScienceFramework, Steinacker2022_DemandDrivenDesign}, or highly detailed optimization models explicitly accounting for construction costs and health benefits of cycling \cite{Paulsen2024_WelfareOptimalExpansionInducedDemand, Steinacker2025_RobustNetworkDesign}. However, most planning approaches rely exclusively on the evaluation of the impact of single links to prioritize network expansions. Figure~\ref{fig:FIG2_HamburgFirstOrder} illustrates the potential challenge resulting from this focus on individual link importance in our example setting: distributed demand leads to complex flows of cyclists across the network, with cyclists preferring short trips but avoiding busy streets without dedicated bike infrastructure (Fig.~\ref{fig:FIG2_HamburgFirstOrder}a). Consequently, evaluating the benefits of improving individual links results in an equally complex distribution of link importance across the network (Fig.~\ref{fig:FIG2_HamburgFirstOrder}b). While some important corridors for cyclists may be visible, planning network expansions based only on this first-order importance of individual links results in scattered bike paths and a disconnected cycling network (Fig.~\ref{fig:FIG2_HamburgFirstOrder}c). In the following, we explain how accounting for higher-order synergies between different link upgrades overcomes this problem.

\vspace{3mm}

\clearpage

\section*{Results}

Consider a generic setting with the transportation infrastructure represented as a weighted network $G = (V,E,\left\{u_e\right\})$, consisting of directed links $E$ (e.g., streets or paths) between nodes $V$ (e.g., intersections or points of interest) with weights $\left\{u_e\right\}$ for each link $e \in E$, representing the link quality (including, for example, travel times, costs, or reliability). The quality of the infrastructure depends on how efficiently it enables all users to travel across the network, described by the flows $X$ for all trips $(o,d) \in \Omega \subseteq V \times V$ from origins $o \in V$ to destinations $d \in V$ summarized in the trip set $\Omega$. Here, the network usage $X = \left(x_{od_1,e_1}, x_{od_1,e_2}, \ldots x_{od_1,e_{\left|E\right|}}, x_{od_2,e_1} \ldots x_{od_{\left|\Omega\right|},e_{\left|E\right|}}\right)^\mathrm{T}$ separately tracks the flows for each origin-destination pair traveling across each directed link in the network in a vector with length $\left|E\right| \times \left|\Omega\right|$. The overall network performance, describing, for example, the total travel time, is given by a utility function $U(G, X)$, typically as the sum of the utilities $U_{od}(G, X)$ for all individual trips,
\begin{equation}
    U(G, X) = \sum_{(o,d) \in \Omega} U_{od}(G, X) \,. \label{eq:overall_utility}
\end{equation}

\subsection*{Quantifying link importance and synergies}

To systematically understand the impact of individual links and the benefits of jointly improving multiple links, we evaluate upgrades of the existing infrastructure such that the structure of the network $G$, i.e. the set of nodes $V$ and links $E$, remains unchanged. Only the quality $u_e$ of a specific set of links $e \in \Delta E \subseteq E$ is improved (or worsened) by some amount $\Delta u_e$, changing $\left\{u_e\right\}$ to $\left\{u_e^\prime\right\}$. Improving links in the network affects the network performance $U(G,X)$ both directly and indirectly: The explicit dependence on the infrastructure $G$ via the link qualities $\left\{u_e\right\}$ captures the direct effects of the link upgrades; network upgrades also change the usage $X$, in turn indirectly affecting network performance.

We write the performance $U(G^\prime,X^\prime)$ of the improved network $G^\prime = (V,E,\left\{u_e^\prime\right\})$ with new flows $X^\prime$ reflecting the route choice changes of the users as a series expansion
\begin{equation}
    U(G^\prime,X^\prime) =
    U(G,X) +
    \underbrace{\sum_{e \in \Delta E} \frac{\mathrm{d} U}{\mathrm{d} u_e}\, \Delta u_e }_{I}
    + \underbrace{\frac{1}{2} \sum_{e \in \Delta E} \frac{\mathrm{d}^2 U}{\mathrm{d} u_e^2}\, \Delta u_e^2 }_{S^\mathrm{self}}
    + \underbrace{\frac{1}{2} \sum_{\substack{e_1, e_2 \in \Delta E \\ e_1 \neq e_2}} \frac{\mathrm{d}^2 U}{\mathrm{d} u_{e_1}\,\mathrm{d} u_{e_2}}\, \Delta u_{e_1}\,\Delta u_{e_2} }_{S^\mathrm{cross}}
    + \ldots
    \label{eq:utility_change}
\end{equation}
in powers of the changes in link qualities $\Delta u_e$. Here we have explicitly separated the first-order changes $I$, the second-order effects $S^\mathrm{self}$ of upgrading an individual link, and the second-order terms $S^\mathrm{cross}$ originating from leading higher-order interactions among pairs of links. This expansion is formally correct as long as the total network performance $U(G,X)$ varies smoothly with the link qualities.

The zeroth-order term in the expansion is simply the original network performance $U(G,X)$ in the network without any upgrades. The first-order terms describe the individual link importance
\begin{equation}
    I_e = \frac{\mathrm{d} U}{\mathrm{d} u_e} \, \Delta u_e = \frac{\partial U}{\partial u_e}\, \Delta u_e + \frac{\partial U}{\partial X}\,\frac{\partial X}{\partial u_e}\, \Delta u_e = \frac{\partial U}{\partial u_e} \, \Delta u_e + \sum_{(o,d) \in \Omega}\,\sum_{e^\prime \in E} \frac{\partial U}{\partial x_{od,e^\prime}}\,\frac{\partial x_{od,e^\prime}}{\partial u_e} \, \Delta u_e \label{eq:first_derivative_linkimportance}
\end{equation}
and capture the direct effects of the network upgrades $\frac{\partial U}{\partial u_e}$ for the original flow $X$ as well as effects of route changes $\frac{\partial X}{\partial u_e}$ induced by the upgrade but measured with respect to the original network performance, $\frac{\partial U}{\partial X}$. For ease of notation, in the following we write the chain rule derivatives with respect to the flow $X$ only in short form.

The different second-order terms capture synergies in the network. Second derivatives with respect to the same link upgrade define self-synergies
\begin{equation}
    S_e^\mathrm{self} = \frac{1}{2}\,\frac{\mathrm{d}^2 U}{\mathrm{d} u_e^2} \, \Delta u_{e}^2  = \frac{1}{2}\,\Bigg[\;\frac{\partial^2 U}{\partial u_e^2} + \left[\frac{\partial^2 U}{\partial X^2}\,\left(\frac{\partial X}{\partial u_e}\right)^2 +\frac{\partial U}{\partial X}\,\frac{\partial^2 X}{\partial u_e^2}\right] + 2 \, \frac{\partial}{\partial X}\left[\frac{\partial U}{\partial u_e}\right]\,\frac{\partial X}{\partial u_e} \;\Bigg]\, \Delta u_{e}^2 \,,
    \label{eq:second_derivative_expansion_selfsynergy}
\end{equation}
largely consisting of second-order corrections to terms already present in the first-order link importance, evaluating the direct impact of the link upgrade or the effect of route choices with respect to the original link qualities. Additionally, the last term in Eq.~\ref{eq:second_derivative_expansion_selfsynergy} captures the interactions between network structure and usage by jointly evaluating the effect of flow changes $\frac{\partial X}{\partial u_e}$ with respect to the improvements of the link quality $\frac{\partial}{\partial X}\left[\frac{\partial U}{\partial u_e}\right]$, in contrast to the separate evaluation in the first-order contributions. This effectively describes self-synergies of the link upgrade with the induced change in network usage, such as more flow across the improved link increasing the benefit of the link upgrade.

Finally, second-order derivatives with respect to different links capture cross-synergies
\begin{eqnarray}
    S_{\{e_1, e_2\}}^\mathrm{cross} &=& \frac{\mathrm{d}^2 U}{\mathrm{d} u_{e_1}\,\mathrm{d} u_{e_2}} \, \Delta u_{e_1} \, \Delta u_{e_2} \label{eq:second_derivative_expansion_crosssynergy}\\
    &=& \Bigg[ \phantom{+} \frac{\partial^2 U}{\partial u_{e_1} \, \partial u_{e_2}} + \left[\frac{\partial^2 U}{\partial X^2}\,\frac{\partial X}{\partial u_{e_1}}\,\frac{\partial X}{\partial u_{e_2}} +\frac{\partial U}{\partial X}\,\frac{\partial^2 X}{\partial u_{e_1} \, \partial u_{e_2}}\right] \nonumber \\
    && \phantom{\Bigg[} + \left[\left(\frac{\partial}{\partial X} \left[\frac{\partial U}{\partial u_{e_1}}\right]\right)\,\frac{\partial X}{\partial u_{e_2}} + \left(\frac{\partial}{\partial X} \left[\frac{\partial U}{\partial u_{e_2}}\right]\right)\,\frac{\partial X}{\partial u_{e_1}}\right] \, \Bigg] \, \Delta u_{e_1} \, \Delta u_{e_2}\nonumber
\end{eqnarray}
of upgrading both links $e_1$ and $e_2$ together (compare Fig.~\ref{fig:FIG1_SchematicNetworkImprovement}). These represent genuine higher-order interactions between the two links. The first term captures direct dependencies of the network performance on both link qualities. Most common traffic models do not include such effects, though they become relevant, for example, when explicitly modeling the discomfort cyclists experience when repeatedly leaving a dedicated bike path and merging with car traffic \cite{Teschke2012_RouteInfrastructureRisk}. The second term extends the dependence of the network performance on the flow changes already present in the first-order contribution to second-order effects of the joint link upgrades. Finally, the third term describes the benefit $\frac{\partial}{\partial X}\left[\frac{\partial U}{\partial u_{e_1}}\right]$ of one link upgrade for the change $\frac{\partial X}{\partial u_{e_2}}$ in network usage induced by the other link upgrade and vice versa. This contribution captures cross-synergies between two links not accounted for in the terms focusing on the impact of a single link.

For larger sets $\Delta E$ upgrading three or more links, we define the second-order cross-synergy of the set as the sum of all pairwise cross-synergies between links in the set,
\begin{equation}
    S_{\Delta E}^\mathrm{cross} = \sum_{\substack{\{e_1,e_2\} \subset \Delta E\\ e_1 \neq e_2}} S_{\{e_1,e_2\}}^\mathrm{cross} \,. \label{eq:cross_synergies_multiple}
\end{equation}
The benefits of jointly upgrading two sets $\Delta E_1$ and $\Delta E_2$ of links are described by their cross-synergy benefit
\begin{equation}
    \Delta S_{\Delta E_1, \Delta E_2}^\mathrm{cross} = S_{\Delta E_1 \cup \Delta E_2}^\mathrm{cross} - \left[S_{\Delta E_1}^\mathrm{cross} + S_{\Delta E_2}^\mathrm{cross} \right] \label{eq:cross_synergy_benefit}
\end{equation}
gained (or lost) compared to upgrading the links in the sets $\Delta E_1$ and $\Delta E_2$ individually.
Generally, higher-order terms of the expansions could also be interpreted as higher-order cross-synergies over larger sets of links; however, their calculation quickly becomes prohibitively costly.

For small improvements $\Delta u_e$, the individual second-order terms in Eq.~\ref{eq:utility_change} are often small compared to the first-order terms. However, when upgrading multiple links, the number of second-order synergy terms increases quadratically with the number $\left|\Delta E\right|$ of upgraded links, whereas the number of first-order link importance terms grows only linearly. Thus, the cross-synergies may substantially contribute to the total change of network performance when upgrading multiple links in a network even if individual terms $S_{\{e_1, e_2\}}^\mathrm{cross}$ are small.

\newpage

\subsection*{Evaluating link importance and synergies via cycle flows}

To evaluate link importance and synergies, we need to evaluate all terms contributing to the expressions in Eq.~\eqref{eq:first_derivative_linkimportance}-\eqref{eq:second_derivative_expansion_crosssynergy}, requiring both a model for the total network performance $U$ and for the network flows $X$. For a given model of the network performance $U$, the partial derivatives of $U$ are known functions evaluated at the original state with the network $G$ and the flows $X$. However, the impact of the network upgrades on the flows, $\frac{\partial X}{\partial u_e}$, is typically more challenging to compute. Explicitly sampling all possible link upgrades and recomputing flow changes is computationally unfeasible. For our example network of Hamburg with about $\num{7000}$ links, we would already have to compute the flow changes for almost $\num{24.5}$ million link pairs.

Instead, representing network usage $X$ as the solution of an optimal flow problem,
\begin{equation}
    X = \argmax_{Y} \; C(G,Y) \quad\quad\quad \mathrm{s. t.} \quad Y \succeq 0 \quad\mathrm{and}\quad B\,Y = S \,,\label{eq:optimal_flow}
\end{equation}
provides an alternative, analytical way to track the flow changes. Here, $C(G,X)$ is a typically concave goal function that determines the unique optimal flow in the network. Common examples are socially optimal routing with $C = U$ representing the (negative) total travel time in the network or standard traffic assignment problems where $C$ represents the (negative) cumulative travel time \cite{Roughgarden2002_BadSelfishRouting, Youn2008_PriceanArchyTransportation}. The inequality constraints enforce that all components of the network usage $Y$ are non-negative, ensuring non-negative flows on all directed links in the network. The equality constraints ensure flow conservation in the network for each origin-destination pair, respectively, with $B$ denoting the incidence matrix of the network and $S$ the source-sink vector for travel demand, both split across the different origin-destination pairs.

This representation of the network usage enables us to track the optimal solution as a function of the network upgrades and evaluate the derivatives. To further simplify these calculations, we express the flow changes as cycle flows $X_o$ in the network for each origin-destination pair such that the new flows are $X^\prime = X + X_o$. Since cycle flows have no additional sources or sinks, any flow $X^\prime$ automatically fulfills the original flow conservation constraints like the original flows $X$. For small changes of the link qualities $u_e$, we assume that the set of links with non-zero flow in the network does not change, enabling us to evaluate the cycle flows $X_o$ by considering only the active links with $x_{od,e} > 0$. Since the non-negative flow constraint is not active for these optimal cycle flows $X_o$, they obey the standard optimization condition $\frac{\partial C}{\partial X_o} = 0$. The cycle flows on active links induced by a network upgrade of link $e$ are therefore implicitly defined by
\begin{equation}
    \frac{\mathrm{d}}{\mathrm{d}u_e}\,\left[\frac{\partial C}{\partial X_o}\right] = \frac{\partial^2 C}{\partial u_e\,\partial X_o} + \sum_{o^\prime} \frac{\partial^2 C}{\partial X_{o^\prime}\,\partial X_o}\,\frac{\partial X_{o^\prime}}{\partial u_e} = 0 \,. \label{eq:first_order_cycle_flow_changes}
\end{equation}
Similar to the derivatives of the network performance $U$, the derivatives for a given route choice model with cost function $C$ are known, and we obtain the derivatives $\frac{\partial X_{o^\prime}}{\partial u_e}$ of the cycle flows as the solution of this linear system of equations.

Converting the derivatives of the cycle flows back to derivatives of link flows then enables us to efficiently compute the utility changes in Eq.~\ref{eq:utility_change}. Repeating the same calculations for second derivatives gives a similar set of linear equations that depend on the first derivatives (see Supplementary Note 4 for the full derivation). The theoretical framework and evaluations we proposed and evaluated above equally apply to three-link and other higher-order interactions. However, the computational complexity of evaluating all higher-order derivatives increases exponentially with the order of the interactions. As we demonstrate below, taking second-order terms is sufficient for an accurate approximation of the total network performance.

Supplementary Note 5 provides the full calculations of these derivatives for our example setting with the perturbed utility route choice model. Supplementary Notes 6 explicitly demonstrates the calculations of the derivatives and the usage of the cycle basis in a small example network.

\newpage

\subsection*{Cross-synergies predict network performance}

\begin{figure}[!h]
    \centering
    \includegraphics[width=\linewidth]{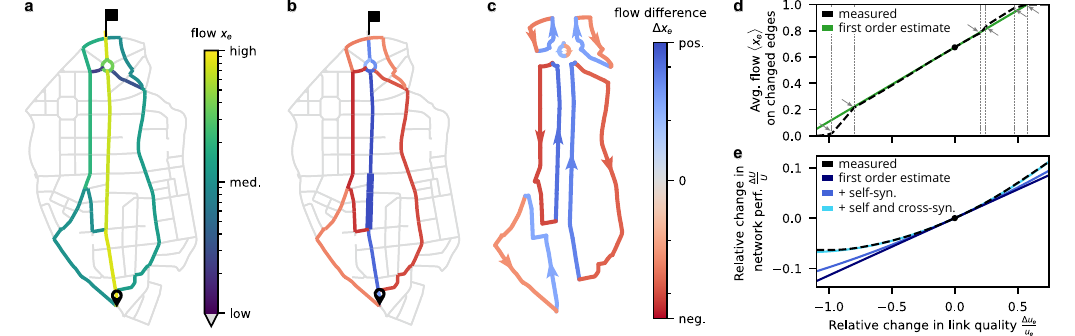}
    \caption{
    \textbf{Cross-synergies accurately predict total network performance changes.}
    (a) Flows between a single origin (black pin) and destination (black flag) in a small example network. The perturbed utility route choice model assigns high flow to the most direct route (yellow and light green), distributes some flow to alternative parallel routes (dark green and blue), but leaves most links in the network unused with zero flow (gray).
    (b) Upgrading a set of four consecutive links in the center of the network (thick blue line) increases the flow on the main route (blue) but decreases the flow on the parallel alternative routes (red).
    (c) These flow changes are composed of $6$ cycle flows.
    (d) For small link utility changes around the original state (circle), the flow changes smoothly. For large link utility changes, the different links become active, and flow changes are non-analytic (arrows). Despite these kinks, our flow estimation (first order, green) still relatively accurately tracks the real flow changes (black dashed).
    (e) Approximations of the network performance based only on the benefits of the individual links (first-order, dark blue, and first-order and self-synergies, blue) are accurate only for small changes close to the original network state (circle) but strongly underestimate the actual network performance (black dashed) for larger deviations. In contrast, including cross-synergies [Eq.~\eqref{eq:cross_synergies_multiple}, light blue] between the upgraded links provides an accurate prediction across the whole range of link quality changes.
    }
    \label{fig:FIG3_utility_estimation}
\end{figure}

To demonstrate the quantitative framework and the perturbed utility route choice model we employ later with large-scale data, to validate the approximations introduced, and to illustrate the interpretation of the network synergies, we first consider a simpler example with a single origin-destination pair (Fig.~\ref{fig:FIG3_utility_estimation}).

The perturbed utility route choice model \cite{Fosgerau2022_PerturbedUtilityRoute, Fosgerau2023_BikeabilityInducedDemand, Fosgerau2026_Sensitivity} constitutes an optimal flow model of the form of Eq.~\eqref{eq:optimal_flow}. The goal function $C$ is a combination of a linear (in the flow $X$) utility, assigning the majority of the flow along the weighted shortest path between origin and destination, and a nonlinear entropy-like perturbation, distributing flow across alternative paths in the network (see Methods for details). However, instead of distributing flow across all available links, the form of the perturbation ensures that most links in the network remain unused, closely reflecting empirical route choice behavior (Fig.~\ref{fig:FIG3_utility_estimation}a).

Improving a set $\Delta E$ of links on the main route increases the flow along this path and decreases the flow along alternative routes (Fig.~\ref{fig:FIG3_utility_estimation}b). Due to the structure of the active links, these flow changes are described by only $6$ different cycle flows (Fig.~\ref{fig:FIG3_utility_estimation}c). For small changes of the link utilities, the flows change smoothly (Fig.~\ref{fig:FIG3_utility_estimation}d). However, for larger changes, alternative routes become entirely unused, or new routing options are established, resulting in non-analytic kinks in the flow across the upgraded links. While these non-analytic flow changes formally invalidate our series expansion [Eqns.~\eqref{eq:utility_change} and \eqref{eq:first_order_cycle_flow_changes}], we observe that our predictions remain accurate even for large link utility changes.

The predicted flow changes translate into an accurate prediction of the overall network performance $U = C$ (see Methods for details). By explicitly including the cross-synergies, our second-order estimate Eq.~\eqref{eq:utility_change} accurately predicts the total network performance over the whole range of link utility changes (Fig.~\ref{fig:FIG3_utility_estimation}e). In contrast, first-order estimates of the network performance changes or even the self-synergies of the upgraded links do not capture the redistribution of flows and strongly underestimate the actual utility benefits (and equivalently overestimate the utility penalties when decreasing link utilities).

\vspace{3mm}

\newpage

\subsection*{Synergistic network structures}

\begin{figure}[!h]
    \centering
    \includegraphics[width=\linewidth]{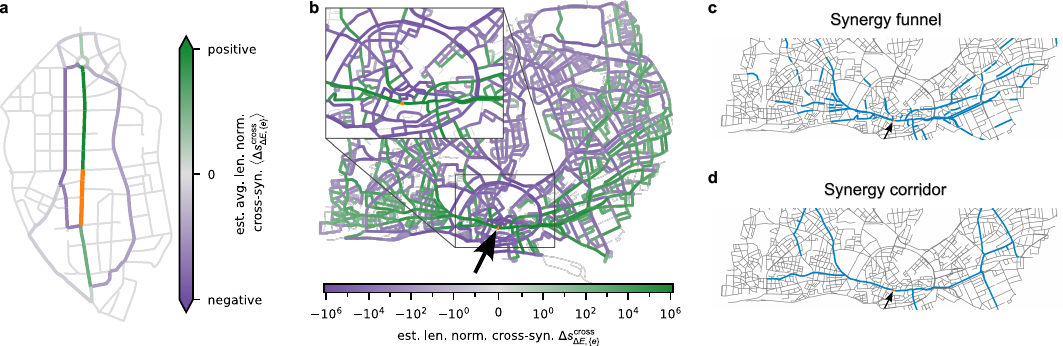}
    \caption{
    \textbf{Synergies promote connected network structures.}
    (a) Cross-synergies [Eq.~\ref{eq:second_derivative_expansion_crosssynergy}] quantify how upgrading one link affects the benefits of upgrading another link. In the small example network (compare Fig.~\ref{fig:FIG3_utility_estimation}), upgrading a set of focal links on the main route (thick orange) results in more flow along the main route and therefore increases the benefits of also upgrading the remaining links on that route (positive cross-synergy, green), since users benefit from both upgrades. Simultaneously, it decreases the benefits of upgrading links on parallel routes (negative cross-synergy, purple), since each user can only benefit from one of the upgrades.
    (b) The same general pattern also emerges for link upgrades in more complex networks (Hamburg, compare Fig.~\ref{fig:FIG1_SchematicNetworkImprovement}a-c). When improving a single focal link (orange, marked by the arrow), other links on the same routes (in series) have a positive cross-synergy (green), whereas links on alternative routes (in parallel) tend to have negative cross-synergy (purple).
    (c) The $150$ links (blue) with the highest length-normalized cross-synergy with the focal link (orange) form a tree-like network, funneling flow towards the focal link. However, since the absolute value of the cross-synergies quickly decreases with distance, the reach of this network remains relatively small.
    (d) We grow an extended network by iteratively adding the link with the highest normalized cross-synergy with all previous links. This process results in a connected corridor of $150$ synergistic links (blue) funneling east-west flow across the focal link. Since adjacent links often carry similar flow and therefore have a high cross-synergy, this corridor is naturally connected.
    \vspace{-30mm}
    }
    \label{fig:FIG4_Synergy_structure}
\end{figure}

The cross-synergies defined in Eq.~\ref{eq:second_derivative_expansion_crosssynergy} describe which sets of links benefit from joint upgrades. In our small example setting, computing the cross-synergies of each link $e$ with the set $\Delta E$ of upgraded links along the direct route naturally highlights other links along the same path (Fig.~\ref{fig:FIG4_Synergy_structure}a; compare Fig.~\ref{fig:FIG3_utility_estimation}). Here, the set $\Delta E$ acts as a focal set around which we evaluate the added synergies of individual links [compare Eq.~\eqref{eq:cross_synergy_benefit}]. To avoid a bias towards longer links with a larger impact on the route, we consider normalized cross-synergy benefits
\begin{equation}
    \Delta s_{\Delta E, \{e\}}^\mathrm{cross} = \frac{\Delta S_{\Delta E , \{e\}}^\mathrm{cross}}{l_e} = \frac{1}{l_e} \, \sum_{e^\prime \in \Delta E} S_{\{e,e^\prime\}}^\mathrm{cross}  \label{eq:length_normalized_crosssynergy_benefit}
\end{equation}
relative to the length of the additional link. The most synergistic upgrades are those links directly adjacent to the focal set that carry the same flow, since the increased flow induced by the focal link upgrades would also benefit from upgrades of these adjacent links. In contrast, upgrading links on the alternative connections is anti-synergistic since flow would be diverted away from the upgraded focal links, represented by a negative cross-synergy.

The same qualitative features of link synergies are also visible in the full bicycle infrastructure network of Hamburg with its more complex demand structure with respect to a single bidirectional focal link $\Delta E = \{e_{ij}, e_{ji}\}$ (Fig.~\ref{fig:FIG4_Synergy_structure}b). Here and in the following, we jointly consider both forward and backward directions of each link by adding the importance and synergies of the links in both directions to keep the visualizations interpretable. Positive cross-synergies highlight the set of links that funnel flow towards the focal link to maximize the benefits of upgrading it. Conversely, parallel connections or links routing flow perpendicular to the focal link are anti-synergistic. In general, the absolute value of the (anti)synergy decreases with increasing distance, since the flow distributes across more links further away from the focal link (see Supplementary Note 7 for additional examples and Supplementary Note 8 for a brief analysis of the distance scaling).

In contrast to upgrading the links with the highest individual importance, the set of most synergistic links forms an almost connected, tree-like structure centered around the focal link (Fig.~\ref{fig:FIG4_Synergy_structure}c). However, this set still contains many parallel links. Adopting a dynamic approach, we grow a synergistic set $\Delta E^*$ of links from the initial bidirectional focal link by iteratively adding the bidirectional link with the highest length-normalized cross-synergy benefit with all previously added links (see Methods for details). Due to the cross-synergies increasing with shared flow, the links added during this process self-organize into a mostly connected network growing outward from the focal link while avoiding parallel connections. The synergistic links effectively form a branching corridor for all trips in the east-west direction aligned with the focal link (Fig.~\ref{fig:FIG4_Synergy_structure}d).

\vspace{3mm}

\clearpage

\subsection*{Identifying synergistic network backbones}

\begin{figure}[!ht]
    \centering
    \includegraphics[width=\linewidth]{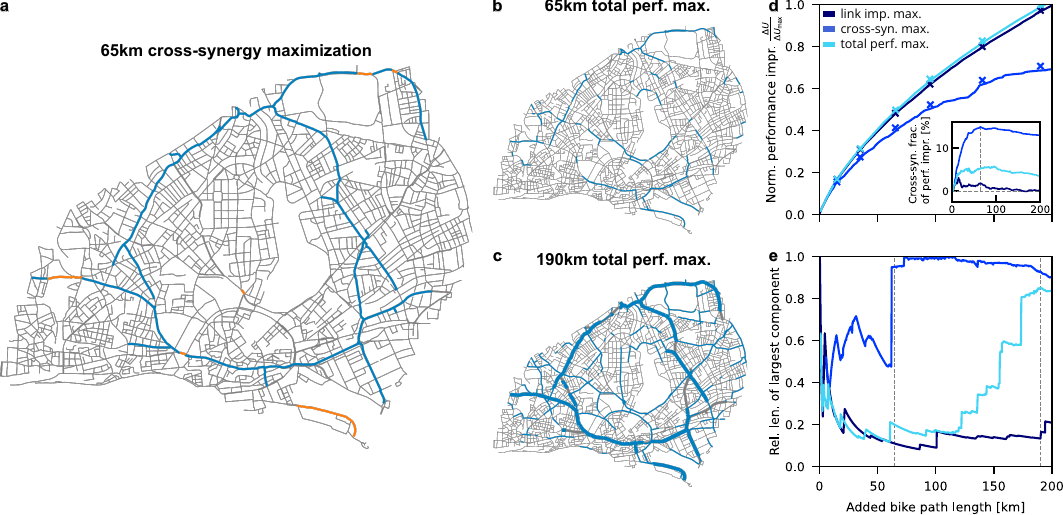}
    \caption{
    \textbf{Synergies define the backbone of connected cycling infrastructure.}
    (a) Starting from the set of the ten bidirectional links with the highest length-normalized first-order importance (orange), we iteratively add links with the largest length-normalized cross-synergy (see Methods for details). The upgraded links (blue) form a connected network backbone already for a small total length of $\SI{65}{\km}$ of added bike paths, maximizing the contributions of cross-synergies to total network performance (compare panel d inset).
    (b,c) In contrast, optimizing for the total utility up to and including second-order terms results in a set of scattered upgrades at the same total length. The network only becomes connected when adding $\SI{190}{\km}$ of bike paths. It then contains a large part of the synergy backbone (thick blue links; compare panel a, gray links show parts of the backbone not included in the network).
    (d) Optimizing for total network performance up to first- and second-order contributions (dark and light blue, respectively) achieves a larger overall benefit (normalized to the maximum benefit with $\SI{190}{\km}$ upgraded bike paths) than optimizing cross-synergies (blue). However, optimizing for cross-synergies maximizes these second-order contributions for $\SI{65}{\km}$ of added bike paths (see inset, compare panel a), thereby promoting network connectivity (compare panel e). In all cases, the predicted network performance changes (lines) accurately estimate actual improvements (crosses, see also Supplementary Note 9).
    (e) Connectivity of the bike path network measured as the fraction of the largest connected component relative to the total length of the upgraded links. Dashed vertical lines mark $\SI{65}{\km}$ and $\SI{190}{\km}$ of upgraded bike paths as illustrated in panels a-c. Optimizing cross-synergies quickly results in a connected backbone structure (blue, compare panel a). Optimizing for the total utility up to second-order terms eventually results in a connected network (light blue), whereas considering only first-order individual link importance (dark blue) leaves the upgraded links entirely disconnected (compare Fig.~\ref{fig:FIG2_HamburgFirstOrder}c).
    \vspace{-25mm}
    }
    \label{fig:FIG5_Synergy_backbone}
\end{figure}

We exploit these features of cross-synergies to identify a connected backbone of synergistic links for the entire network (Fig.~\ref{fig:FIG5_Synergy_backbone}). Instead of starting with a single focal link, we take the $10$ bidirectional links with the highest length-normalized first-order importance as our initial set $\Delta E^*_0$, seeding the backbone throughout the network. We then iteratively add links with the largest length-normalized cross-synergy as above. The resulting set of links quickly grows into a connected ring-like structure covering the entire the network (Fig.~\ref{fig:FIG5_Synergy_backbone}a) Adding further links creates a denser mesh of network upgrades (see Supplementary Note 9).

At the same total length of bike paths, direct optimization of the total network performance, including all first- and second-order length-normalized contributions to the total utility (see Methods for details), results in a highly fragmented network of infrastructure upgrades (Fig.~\ref{fig:FIG5_Synergy_backbone}b). Only when adding further upgrades does the network eventually become connected, containing a large part of the synergy backbone (Fig.~\ref{fig:FIG5_Synergy_backbone}c). In contrast, optimizing solely based on the first-order contribution of individual links captures a large part of the possible network performance improvements, but the resulting network upgrades remain disconnected (Fig.~\ref{fig:FIG5_Synergy_backbone}d,e and compare Fig.~\ref{fig:FIG2_HamburgFirstOrder}c). While optimizing for cross-synergies sacrifices some of the total network performance, the large contribution of cross-synergies (Fig.~\ref{fig:FIG5_Synergy_backbone}d, inset) ensures connectivity. This contribution would be entirely absent or even negative due to anti-synergies with first-order optimization. Including the second-order contributions to the total network performance is thus crucial to obtain connected rather than scattered infrastructure upgrades(Fig.~\ref{fig:FIG5_Synergy_backbone}e).

See Supplementary Note 9 for additional snapshots of the network upgrades and quantitative evaluations of the first- and second-order contributions. Supplementary Note 10 illustrates similar results on the street network of Manhattan, demonstrating the robustness of our approach.

\vspace{3mm}

\section*{Discussion}

Higher-order interactions emerge naturally in transport infrastructure networks where trips across the network inherently span multiple links \cite{Xu2016_HigherorderDependenciesNetworks, Lambiotte2019_NetworksoHigherorderModels, Pappalardo2023_FutureDirectionsMobility, Kim2024_ShortestPathPercolation, Kim2025_ModelingResourceConsumption}. While we illustrated our approach for bicycle traffic with a specific route choice model, the mathematical framework we introduced is directly applicable to other transport networks where usage is described by the solution of an optimal flow problem. Supplementary Note 11 illustrates this generality by considering standard traffic assignment with congestion in a small example network. While the spatial structure of the (anti)synergies depends on the details of the interactions between different origin-destination flows, understanding these variations may highlight essential differences of efficient network structures for different transport modes. This generic applicability may also enable a unified analysis of higher-order interactions and (anti)synergies in multi-modal transport, such as converting car lanes into bike infrastructure in combined car and bike traffic sharing limited urban space \cite{Goessling2016_UrbanSpaceDistribution, Wiedemann2025_BikeNetworkPlanning, Creutzig2020_FairStreetSpace}. Moreover, the general concept of higher-order synergies may be applicable beyond transport infrastructure in arbitrary flow networks such as electric power grids or water and gas supply networks \cite{Schaefer2022_UnderstandingBraessParadox, Xu2016_HigherorderDependenciesNetworks}.

We have defined synergies between different links as second-order terms in a series expansion of the total network performance. Formally, our approach is thus only correct for small changes to the network where the series expansion Eq.~\eqref{eq:utility_change} remains valid. Despite this formal mathematical constraint, our results demonstrate that the second-order expansion accurately predicts network performance, even when network usage changes non-smoothly, and provides reasonable predictions also for large changes of the network when upgrading many links. However, these predictions still rely on continuous changes of the network usage to obtain meaningful derivatives to evaluate flow changes and their impact on the overall network performance. Extending the concept of synergies to larger structural changes of the underlying network would require the evaluation of non-local flow differences instead of local derivatives, likely requiring explicit and computationally expensive evaluation of each potential structural change. As previously applied in first-order optimization \cite{Steinacker2022_DemandDrivenDesign, Steinacker2025_RobustNetworkDesign, Paulsen2024_WelfareOptimalExpansionInducedDemand}, a dynamic approach recomputing the network usage and derivatives after each upgrade may help to include such network extensions and further increase the accuracy of our predictions for large changes combining many network upgrades.

Overall, taking the general perspective of higher-order interactions, our results offer a systematic quantitative description of the mutual influences between network structure and usage that enables us to identify synergistic network backbones. Tracking induced changes of the network usage in the cycle basis of the underlying infrastructure network, we have developed a scalable framework to efficiently evaluate first-order link importance and higher-order synergies even in large transport networks with a complex demand distribution. We have exemplified our results for bicycle traffic in the city of Hamburg, Germany, proposing efficient network upgrades that capture higher-order synergies and self-organize into a connected backbone of the infrastructure network. Exploiting these synergies is crucial for unlocking the full potential of infrastructure networks \cite{Pucher2008_MakingCyclingIrresistible, Creutzig2016_UrbanInfrastructureClimate}. The advantage of generating connected network upgrades is especially relevant for cycling infrastructure, where disconnected networks require frequent merging in and out of car traffic, strongly increasing the risk of accidents \cite{Teschke2012_RouteInfrastructureRisk}. Our results may thus improve on previous planning tools that rely solely on first-order link importance measures \cite{Steinacker2025_RobustNetworkDesign, Szell2022_GrowingUrbanBicycle} and provide guidelines for planning mutually beneficial network upgrades and expansions \cite{Olmos2020_DataScienceFramework, Pucher2008_MakingCyclingIrresistible, Sebastiao2026_TradeDirectnessCoverage}. More generally, the presented approach may improve our theoretical understanding of the structural properties of efficient transport networks \cite{Barthelemy2011_SpatialNetworks, Verma2016_EmergenceCorePeripheries, Gastner2006_OptimalSpatialDistrNetworks} and help explain subtle differences across transport modes caused by higher-order interactions between network structure and usage.

\clearpage

\section*{Methods}

\subsection*{Hamburg street network data}
We obtain the street network data from OpenStreetMap \cite{osm}, taking the car-accessible network without car-only links like highways. We simplify the raw network by consolidating nodes within a radius of less than $\SI{20}{\metre}$ (e.g. simplifying the detailed structure of intersections), and placing the resulting consolidated node at the centroid of their former position \cite{OSMnx}. Additionally, we remove all isolated nodes and self-loops that may be present in the data. To ensure routing is possible between all pairs of nodes, we restrict the remaining network to the largest strongly connected component. After all simplification steps, the graph in the simulation area for Hamburg has $\num{2421}$ nodes and $\num{6992}$ directed links.

To generate the demand model, we initially work with a larger area, including the simulation area (e.g. shown in Fig.~\ref{fig:FIG1_SchematicNetworkImprovement}a) and a buffer zone of \SI{5}{\km} around it. We reduce the city-wide graph to the graph of the simulation area by mapping all trips starting and/or ending outside the simulation area to the nodes where they enter or leave this area (see Supplementary Note 1) and discarding all nodes outside the bounding polygon \cite{OSMnx}. The full demand model is described in detail in Supplementary Note 2.

\subsection*{Perturbed utility route choice}
We model bicycle route choice using the perturbed utility model \cite{Fosgerau2022_PerturbedUtilityRoute, Fosgerau2023_BikeabilityInducedDemand, Fosgerau2026_Sensitivity}, representing the flow $X$ of cyclists on the network $G$ as the solution of an optimal flow problem by maximizing the goal function $C(G,X)$. For simplicity, we take the same function to evaluate the network performance, $U(G,X) = C(G,X)$.

We assume that individual cyclists do not interact (no congestion on bike paths) such that the route choice problem separates into independent individual trips for which we assign unit flow $X_{od}$ from origin $o$ to destination $d$ by maximizing the perturbed utility
\begin{align}
    U_{od}(G,X_{od}) &= \sum_{e \in E} l_e\,u_e\,x_{od,e} - \sum_{e \in E} l_e\,\left[(1+x_{od,e})\,\mathrm{ln}\left(1+x_{od,e}\right) - x_{od,e}\right] \,. \label{eq:purc}
\end{align}
The first sum denotes the (negative) utility for cyclists to travel along a link $e$, taken here as a linear function of the flow $x_{od,e} \ge 0$ with a utility rate $u_e < 0$ per length $l_e \ge 0$ of link $e$. This term favors shortest path routing. The second sum represents the perturbation term motivated by statistical physics entropy, favoring the distribution of demand across different paths but allowing routes and links to remain unused. Consequently, the cyclists predominantly choose the shortest path, but the flow also distributes itself over multiple alternative paths (compare Fig.~\ref{fig:FIG3_utility_estimation}a). In stark contrast to logit models or related statistical physics interpretations, however, most route options and thereby also most links in the network remain unused, $x^*_{od,e} = 0$, reproducing the results of empirical route choice experiments \cite{Fosgerau2022_PerturbedUtilityRoute}.

Evaluating the network performance with the same function optimized in the route choice model, $U = C$, strongly simplifies the derivatives \cite{Fosgerau2026_Sensitivity}. Since the flows optimize $U$, any redistribution of the flow on the active links as a reaction to a network upgrade cannot change the network performance evaluated with the original link qualities. Consequently, the second term of the link importance cancels, $\frac{\partial U}{\partial X}\,\frac{\partial X}{\partial u_e} = 0$, and the link importance
\begin{equation}
    \frac{\mathrm{d} U_{od}}{\mathrm{d} u_e} \, \Delta u_e = \frac{\partial U_{od}}{\partial u_e}\, \Delta u_e + \frac{\partial U_{od}}{\partial X_{od}}\,\frac{\partial X_{od}}{\partial u_e}\, \Delta u_e = \frac{\partial U_{od}}{\partial u_e}\, \Delta u_e = l_e\,x_{od,e}\, \Delta u_e
\end{equation}
is directly proportional to the original flow and the potential link quality improvement. In this form, the link importance is exactly equivalent to standard importance measures used in previous network planning approaches \cite{Olmos2020_DataScienceFramework, Steinacker2022_DemandDrivenDesign}.

Similarly, the terms derived from $\frac{\partial U}{\partial X}\,\frac{\partial X}{\partial u_e}$ in the self- and cross-synergies cancel. Since the utility is linear in the link qualities and does not include direct interactions, this leaves only the last terms of Eqns.~\eqref{eq:second_derivative_expansion_selfsynergy} and \eqref{eq:second_derivative_expansion_crosssynergy}
\begin{equation}
    S_e^\mathrm{self} = \frac{1}{2}\, \frac{\partial}{\partial X}\left[\frac{\partial U}{\partial u_e}\right]\,\frac{\partial X}{\partial u_e} \, \Delta u_{e}^2 = \frac{1}{2}\, l_e \, \frac{\partial x_e}{\partial u_e} \, \Delta u_{e}^2
\end{equation}
and
\begin{eqnarray}
    S_{\{e_1, e_2\}}^\mathrm{cross}
    &=& \frac{1}{2} \, \Bigg[ \left(\frac{\partial}{\partial X} \left[\frac{\partial U}{\partial u_{e_1}}\right]\right)\,\frac{\partial X}{\partial u_{e_2}} + \left(\frac{\partial}{\partial X} \left[\frac{\partial U}{\partial u_{e_2}}\right]\right)\,\frac{\partial X}{\partial u_{e_1}} \, \Bigg] \, \Delta u_{e_1} \, \Delta u_{e_2} \nonumber \\
    &=& \frac{1}{2}\, \left[ l_{e_1} \, \frac{\partial x_{e_1}}{\partial u_{e_2}} + l_{e_2} \, \frac{\partial x_{e_2}}{\partial u_{e_1}} \right] \, \Delta u_{e_1} \, \Delta u_{e_2} \,.
\end{eqnarray}
The second-order terms thus directly capture the interactions between network structure and usage when upgrading one or more links. The two terms in the cross-synergy represent the different orders of the second derivative,
\begin{equation}
    \frac{\partial U^2}{\partial u_{e_2} \, \partial u_{e_1}} = l_{e_1} \, \frac{\partial x_{e_1}}{\partial u_{e_2}} = l_{e_2} \, \frac{\partial x_{e_2}}{\partial u_{e_1}} = \frac{\partial U^2}{\partial u_{e_1} \, \partial u_{e_2}} \,,
\end{equation}
and we included both terms above to make the symmetry of the cross-synergy explicit. Length-normalized synergy benefits are slightly easier to interpret using only one term, for example when adding link $e_2$ to an already upgraded link $e_1$, as
\begin{eqnarray}
    \frac{\Delta S_{\{e_1\}, \{e_2\}}^\mathrm{cross}}{l_{e_2}} = \frac{1}{2\,l_{e_2}}\, \left[ l_{e_1} \, \frac{\partial x_{e_1}}{\partial u_{e_2}} + l_{e_2} \, \frac{\partial x_{e_2}}{\partial u_{e_1}} \right] \, \Delta u_{e_1} \, \Delta u_{e_2} = \frac{\partial x_{e_2}}{\partial u_{e_1}} \, \Delta u_{e_1} \, \Delta u_{e_2} \,.
\end{eqnarray}

The detailed parameters of the perturbed utility route choice model for Hamburg are provided in Supplementary Note 3. Explicit calculations of the first and second derivatives of the flow required to evaluate general network performance derivatives are provided in Supplementary Note 5. The choice $U = C$ thus simplifies the calculations such that only first derivatives of the flows are actually needed.

\subsection*{Network expansion models}
We construct sets of upgraded links $\Delta E^*$ by optimizing different contributions to the total network performance for the street network of Hamburg. As described in the main text, we always consider bidirectional links and add forward and backward pairs $\{e_{ij}, e_{ji}\}$ of directed links, applying the same length-normalization as introduced in Eq.~\eqref{eq:length_normalized_crosssynergy_benefit} for all optimization approaches.

\paragraph*{First-order optimization}
The simplest expansion model starts from an initially empty set $\Delta E^*_0 = \emptyset$ and iteratively adds new bidirectional links $\delta E^* = \{e_{ij}^*, e_{ji}^*\}$ to the upgraded bike path network $\Delta E^*$ purely based on their length-normalized first-order contribution
\begin{align}
    \delta E^* = \argmax_{ \{e_{ij}, e_{ji}\} \subset E \setminus \Delta E}  \frac{I_{e_{ij}} + I_{e_{ji}}}{l_e} \,.
\end{align}
where the length of the links is the same in both directions, $l_e = l_{e_{ij}} = l_{e_{ji}}$.

\paragraph*{Cross-synergy optimization}
The cross-synergy optimization approach starts from an initial set $\Delta E^*_0$ with the ten bidirectional links with the highest length-normalized first-order contribution, $\frac{I_{e_{ij}} + I_{e_{ji}}}{l_e}$. It then iteratively adds new bidirectional links $\delta E^* = \{e_{ij}^*, e_{ji}^*\}$ with the highest cross-synergy benefit with respect to all previously upgraded links $\Delta E^*$
\begin{align}
    \delta E^* = \argmax_{ \{e_{ij}, e_{ji}\} \subset E \setminus \Delta E^*}  \frac{S_{\Delta E^*, \{e_{ij}, e_{ji}\}}^\mathrm{cross}}{l_e} \,.
\end{align}

\paragraph*{Second-order optimization}
Maximizing the total network performance up to second order combines the two aforementioned approaches. The approach starts from an initially empty set $\Delta E^*_0 = \emptyset$ and iteratively adds new bidirectional links $\delta E^* = \{e_{ij}^*, e_{ji}^*\}$ that achieve the largest length-normalized network performance improvement
\begin{align}
    \delta E^* = \argmax_{ \{e_{ij}, e_{ji}\} \subset E \setminus \Delta E^*} \frac{I_{e_{ij}} + I_{e_{ji}} + S_{e_{ij}}^\mathrm{self} + S_{e_{ji}}^\mathrm{self} + \Delta S_{\Delta E^*, \{e_{ij}, e_{ji}\}}^\mathrm{cross}}{l_e}\,.
\end{align}

\newpage

\section*{Data availability}
All data required to reproduce the results is available on GitHub (\url{https://github.com/PhysicsOfMobility/BikePathSynergies}), together with the code.
The demand data is extracted from Kontur population data sets for Germany \cite{Kontur2023_PopulationDataDE} and the USA \cite{Kontur2023_PopulationDataUSA}, distributed under Creative Commons Attribution International (CC BY) license. The networks are heavily based on OpenStreetMap (OSM) data \cite{osm} distributed under Open Data Commons Open Database License (ODbL).

\section*{Code availability}

The code for our algorithm and a guide to reproducing the results are available on GitHub (\url{https://github.com/PhysicsOfMobility/BikePathSynergies}) under AGPLv3 license.

\bibliography{bibliography.bib}

\vspace{-3mm}

\section*{Acknowledgments}
C.S. acknowledges support from the Deutsche Bundesstiftung Umwelt (DBU, German Federal Environmental Foundation).
The project was partially funded by the Deutsche Forschungsgemeinschaft (DFG, German Research Foundation), project number 493613373 to M.S.

\vspace{-3mm}

\section*{Author contribution}
M.S. initiated the research.
C.S and M.S. conceived and planned the research.
H.W., C.S. and M.S. developed the mathematical framework.
H.W. and C.S. wrote the code.
C.S. collected and analyzed the empirical data.
C.S. performed and analyzed the simulations supported by M.S.
All authors contributed to interpreting the results and writing the manuscript.

\vspace{-3mm}

\section*{Competing interest}
The authors declare no competing interests.

\clearpage

\renewcommand{\thefigure}{S\arabic{figure}}
\renewcommand{\thetable}{S\arabic{table}}
\renewcommand{\theequation}{S\arabic{equation}}

\section*{Supplementary Note 1: Network and demand setup}
\begin{figure}[!h]
    \centering
    \includegraphics[width=0.95\linewidth]{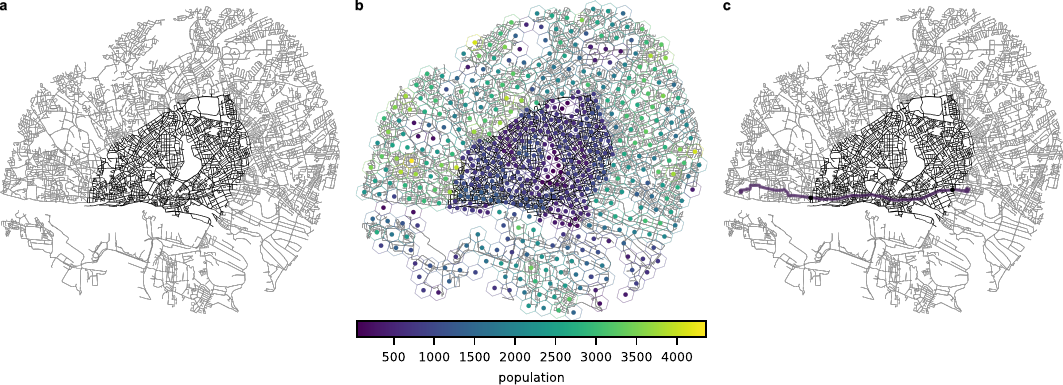}
    \caption{
    \textbf{Network and demand generation for Hamburg.}
    (a) Full network of Hamburg with the simulation area in the center (black) and the buffered area around it (gray).
    (b) Population data based on \cite{Kontur2023_PopulationDataDE} with the color of the cell centroids showing the population number inside the cell. For the buffered area, we keep the original \SI{400}{\metre} hexagon cells. Inside the simulation area, we further divide each cell into equally sized rhombi and split the population accordingly. Due to the smaller cells, the absolute population numbers in the simulation area is lower compared to the buffered area.
    (c) For trips starting and/or ending outside the simulation area (shown for one example, purple) the origin and destination points are moved to the first/last points (black nodes) inside the simulation area along the shortest path.
    \vspace{-20mm}
    }
    \label{fig:SFIG1_graph_demand}
\end{figure}

To generate a synthetic cycling demand in a simulation area, we generate trips based on a singly-constrained gravity model (see Supplementary Note 2 for details) in a larger area and map trips crossing into/out of the simulation area to the first node in the simulation area. As explained in the Methods of the main manuscript, we initially work with a larger area, consisting of the simulation area and a buffer zone of \SI{5}{\km} around it (Figure~\ref{fig:SFIG1_graph_demand}a).

We generate a limited number of origin-destination pairs, based on the Kontur population data set in the \SI{400}{\metre} H3 hexagon resolution \cite{Kontur2023_PopulationDataDE, Kontur2023_PopulationDataUSA} (Figure~\ref{fig:SFIG1_graph_demand}b). To achieve a finer distribution of origin and destination nodes, we split the hexagons inside the simulation area into three equally sized rhombi and distribute the population equally. We then assign the demand to the network nodes closest to the centroids of the cells (hexagons/rhombi) as the origin and destination points for our demand estimation, as long as the nearest node is within \SI{400}{\metre} of the centroid. For Hamburg, this procedure results in $580$ cells in total, with $224$ inside the simulation area and an additional $356$ in the buffer zone. We discard $42$ cells since no network node is sufficiently close to the centroid of the cell (e.g. in the harbor), leaving us with $538$ origin/destination points in the buffered area with approximately $289\,000$ unique trips.

Finally, we map these trips to the simulation area, considering only the shortest physically paths between the respective origin and destination. $32\,000$ trips do not enter the simulation area at all and are therefore discarded. Of the remaining $257\,000$ trips, $208\,000$ trips which start/end outside the simulation area are mapped into the simulation area, by changing their origin/destination to the first/last point on the shortest path that lies within the simulation area (Figure~\ref{fig:SFIG1_graph_demand}c). In this step, multiple trips with different origin/destination points outside the simulation area may be mapped to the same first/last node inside the simulation area. In the end, we obtain approximately $66\,000$ unique trips within the simulation area between $287$ origin/destination points.

\vspace{3mm}

\clearpage

\section*{Supplementary Note 2: Cycling demand model}
Based on the network and population distribution model described in the Methods in the main manuscript and Supplementary Note 1 above, we create the cycling demand model in a two-step process, first estimating total travel demand via a singly constrained gravity model, then computing the fraction of cyclists for each trip as a function of the shortest path distance.

The total number of travelers for a trip from origin $o$ to destination $d$ is given by
\begin{align}\label{eq:30_trip_prob}
    n_{\mathrm{total}}(o,d) = \frac{N(o)\,N(d)}{Z(o)} \, e^{-\frac{L(o, d)}{L_{\mathrm{all}}}} \quad\quad \mathrm{with} \quad \, \sum_d n_{\mathrm{total}}(o,d) = N(o)\,,
\end{align}
where $N(o)$ and $N(d)$ is the population of the origin $o$ and destination $d$, respectively, and $Z(o)$ assures the normalization of the destination distribution for each origin. The trip probability decreases with increasing physical shortest-path distance $L(o,d)$ between the origin and destination in the network (neglecting route choice decisions). We set $L_{\mathrm{all}} = \SI{10}{\kilo\metre}$, denoting the characteristic length scale for trips in the system across all modes, to represent the typical trip length for intra-urban trips \cite{Berrill2024_Comparingurbanform}.

From this general demand $n_{\mathrm{total}}(o,d)$ we extract the demand for cycling by masking short and long trips such that
\begin{align}
    n_{\mathrm{bike}}(o,d) = n_{\mathrm{total}}(o,d) \, \left[ 1 - e^{-\Tilde{p}_{\mathrm{bike}} f(L(o, d)) } \right] \,,
\end{align}
where $\Tilde{p}_{\mathrm{bike}}$ is a parameter to adjust the desired modal split for cyclists and $f(L) \in [0,1]$ captures the distance-dependence of the cycling mode choice, ensuring that bike trips are more likely made for intermediate distances than for very short or long trips \cite{Huber2025_CityCyclingBehaviour}. Here, we take
\begin{align}
    f(L) &= \left( \frac{L}{L_{\mathrm{bike}}} \right)^2 e^{-\frac{L}{L_{\mathrm{bike}}}} \,,
\end{align}
where $L_{\mathrm{bike}} = \SI{2}{\kilo\metre}$ describes the characteristic length scale for bike trips. The peak of $f(L)$, and therefore the highest probability for a trip to be made by bike, is located at $L_{\mathrm{peak}} = \SI{4}{\kilo\metre}$, in line with typical commuting trips by bike \cite{Heinen2011_ChoiceCommuteBike, Lopez2017_UnveilingEBikeCommuting, Plazier2017_EBikeCommuterBehaviour}. Finally, we adjust $\Tilde{p}_{\mathrm{bike}}$, such that the total modal split for bicycles $P_{\mathrm{bike}} = \sum_{o,d} n_{\mathrm{bike}}(o, d) / \sum_o N(o) = 0.2$. Due to the independence of the individual trips in our route choice model, this normalization has no direct impact on our results and only affects the numerical values of the network flows or synergies but not their relative ordering.

For the approximately $66\,000$ unique trips within the simulation area of Hamburg, this results in a total of $90\,000$ cyclists (compare Fig.~1 in the main manuscript).

\section*{Supplementary Note 3: Perturbed utility route choice parameters}
In the main manuscript, we illustrate the approach to quantify synergies in infrastructure networks using a perturbed utility route choice model to describe cyclist route choices as an optimal flow problem \cite{Fosgerau2023_BikeabilityInducedDemand}. The model requires two parameters for each link $e$, the length $l_e$ and the utility rate $u_e$. For the length of a link, we take physical length assigned to the link in the OpenStreetMap data (see Methods in the main manuscript). For the utility rate, we assign each link a value based on its street classification from OpenStreetMap, derived by \citeauthor{Fosgerau2023_BikeabilityInducedDemand} (Tab.~\ref{tab:STAB1_Network_link_categorization})

\begin{table}[!h]
    \centering
    \begin{tabular}{|c|c|c|c|}
         \hline
         Category & $u_{e}$ & $\Delta u_{e}$ & OSM category\\
         \hline\hline
         Base & -0.456 & - & all\\
         \hline
         Bike path & 0.089 & 0.0 & \makecell{track, service, pedestrian, cycleway, path}\\
         \hline
         Residential street & 0.0 & 0.065 & \makecell{everything else}\\
         \hline
         Tertiary street & -0.005 & 0.101 & \makecell{tertiary, tertiary\_link}\\
         \hline
         Secondary street & -0.005 & 0.101 & \makecell{secondary, secondary\_link}\\
         \hline
         Primary street & -0.05 & 0.154 & \makecell{primary, primary\_link}\\
         \hline
    \end{tabular}
    \caption{
        \textbf{Network link categorization.}
        The utility rate $u_e$ as a function of the OpenStreetMap highway tag \cite{Fosgerau2023_BikeabilityInducedDemand}.
        \vspace{-20mm}
        }
    \label{tab:STAB1_Network_link_categorization}
\end{table}

\clearpage

\section*{Supplementary Note 4: Evaluating second order derivatives}
\label{appendix_general_2nd_order}

We evaluate the first-order link importance by computing the first-order derivatives of the flows in the cycle flow basis (see Eq.~(8) in the main manuscript). The same approach enables us to compute the synergies by evaluating second-order derivatives, arriving at a linear system of equations for the second-order derivatives once the first-order derivatives are known.

Computing the second-order derivatives of the optimal flow condition gives
\begin{align}
    0 = \frac{\mathrm{d}}{\mathrm{d}u_{e_2}}\frac{\mathrm{d}}{\mathrm{d}u_{e_1}}\,\left[\frac{\partial C}{\partial X_o}\right] &= \frac{\mathrm{d}}{\mathrm{d}u_{e_2}} \left( \frac{\partial^2 C}{\partial u_{e_1}\,\partial X_o} + \sum_{o^\prime} \frac{\partial^2 C}{\partial X_{o^\prime} \,\partial X_o}\,\frac{\partial X_{o^\prime}}{\partial u_{e_1}} \right) \\
    &= \frac{\partial}{\partial u_{e_2}} \frac{\partial^2 C}{\partial u_{e_1}\,\partial X_o} + \frac{\partial}{\partial u_{e_2}} \sum_{o^\prime} \frac{\partial^2 C}{\partial X_{o^\prime} \,\partial X_o}\,\frac{\partial X_{o^\prime}}{\partial u_{e_1}}  \\
    &\phantom{=} + \sum_{o^{\prime\prime}} \left(\frac{\partial}{\partial u_{o^{\prime\prime}}} \frac{\partial^2 C}{\partial u_{e_1}\,\partial X_o} \right)\frac{\partial X_{o^{\prime\prime}}}{\partial u_{e_2}} + \sum_{o^{\prime\prime}} \left(\frac{\partial}{\partial u_{o^{\prime\prime}}}  \sum_{o^\prime} \frac{\partial^2 C}{\partial X_{o^\prime} \,\partial X_o}\,\frac{\partial X_{o^\prime}}{\partial u_{e_1}} \right)\frac{\partial X_{o^{\prime\prime}}}{\partial u_{e_2}} \nonumber \\
    &= \frac{\partial^3 C}{\partial u_{e_2}\,\partial u_{e_1}\,\partial X_o} + \sum_{o^\prime} \left(\frac{\partial^3 C}{\partial u_{e_2} \, \partial X_{o^\prime} \,\partial X_o}\,\frac{\partial X_{o^\prime}}{\partial u_{e_1}} + \frac{\partial^2 C}{\partial X_{o^\prime} \,\partial X_o}\,\frac{\partial^2 X_{o^\prime}}{\partial u_{e_2} \, \partial u_{e_1}} \right) \\
    &\phantom{=} + \sum_{o^{\prime\prime}} \left( \frac{\partial^3 C}{\partial X_{o^{\prime\prime}} \,\partial u_{e_1}\,\partial X_o} + \frac{\partial^3 C}{\partial X_{o^{\prime\prime}} \,\partial X_{o^{\prime}} \,\partial X_o} \frac{\partial X_{o^\prime}}{\partial u_{e_1}} + \frac{\partial^2 C}{\partial X_{o^{\prime}} \,\partial X_o} \underbrace{\frac{\partial^2 X_{o^\prime}}{\partial X_{o^{\prime\prime}} \,\partial u_{e_1}}}_{=0}\right) \frac{\partial X_{o^{\prime\prime}}}{\partial u_{e_2}} \,, \nonumber
\end{align}
resulting in the linear system of equations
\begin{align}
    & \phantom{=} \sum_{o^{\prime\prime}} \left( \frac{\partial^3 C}{\partial X_{o^{\prime\prime}} \,\partial u_{e_1}\,\partial X_o} + \frac{\partial^3 C}{\partial X_{o^{\prime\prime}} \,\partial X_{o^{\prime}} \,\partial X_o} \frac{\partial X_{o^\prime}}{\partial u_{e_1}} + \frac{\partial^2 C}{\partial X_{o^{\prime}} \,\partial X_o} \right) \frac{\partial X_{o^{\prime\prime}}}{\partial u_{e_2}} \nonumber \\
    &= -\frac{\partial^3 C}{\partial u_{e_2}\,\partial u_{e_1}\,\partial X_o} - \sum_{o^\prime} \left(\frac{\partial^3 C}{\partial u_{e_2} \, \partial X_{o^\prime} \,\partial X_o}\,\frac{\partial X_{o^\prime}}{\partial u_{e_1}} + \frac{\partial^2 C}{\partial X_{o^\prime} \,\partial X_o}\,\frac{\partial^2 X_{o^\prime}}{\partial u_{e_2} \, \partial u_{e_1}} \right) \,,  \label{eq:supp_second_order_cycle_flow_changes}
\end{align}
where again all derivatives of $C$ are known functions and we obtain the second derivatives of the flows $X$ once the first derivatives are known.

\clearpage

\section*{Supplementary Note 5: Detailed calculations\\ for the perturbed utility route choice model}
In the main manuscript and Supplementary Note 4, we presented the calculations of the link importance and synergies for general utility functions. Here, we derive the explicit expressions for the perturbed utility route choice model used to describe cyclist route choices in our main example, following Eqs.~(2-5) in the main manuscript.

The perturbed utility route choice model describes cyclist flows $x_{od,e}$ by maximizing the utility function
\begin{align}
    U_{od}(G,X_{od}) &= \sum_{e \in E} l_e\,u_e\,x_{od,e} - \sum_{e \in E} l_e\,\left[(1+x_{od,e})\,\mathrm{ln}\left(1+x_{od,e}\right) - x_{od,e}\right] \,. \label{eq:S4_purc}
\end{align}
for each trip from origin $o$ to destination $d$ (compare Methods in the main manuscript). Since all trips are independent, we do all calculations for a single trip and omit the $od$ subscript for better readability. The total values of the derivatives, the link importances, and synergies are given by the sum over all trips.

\subsection*{First-order expansion}
The first-order derivative around the optimal flow solution $x^*$, quantifying direct link importances, becomes
\begin{align}
\frac{\mathrm{d} U(X^*)}{\mathrm{d} u_e} &= \frac{\partial U}{\partial u_e} + \sum_{e^{\prime}} \frac{\partial U}{\partial x_{e^{\prime}}^*} \frac{\partial x_{e^{\prime}}^*}{\partial u_e} \\
&= l_e x_e^* + \sum_{e^{\prime}} \big[l_{e^{\prime}} u_{e^{\prime}} - l_{e^{\prime}} \ln(1 + x^*_{e^{\prime}}) \big] \, \frac{\partial x_{e^{\prime}}^*}{\partial u_e} \,.
\end{align}
To evaluate the derivatives $\frac{\partial x_{e^{\prime}}^*}{\partial u_e}$ efficiently, we switch to the cycle basis of the used links as described in the main manuscript. First, we calculate the derivative of the perturbed utility in the cycle basis for a cycle $o$ of used links
\begin{align}
    \frac{\partial U}{\partial x_o} = \sum_{e^{\prime}} b_{e^{\prime} o} \, l_{e^{\prime}} \, u_{e^{\prime}} - l_{e^{\prime}} \, b_{e^{\prime} o} \, \ln \left(1 + x_{e^{\prime}}^* + \sum_{o^{\prime}} b_{e^{\prime} o^{\prime}} \, x_{o^{\prime}}\right) \,.
    \label{eq:delU_delx}
\end{align}
Here, $b_{eo} \in \{-1,0,1\}$ is an indicator function, describing if a link is part of the cycle $o$ in the direction of the flow ($b = 1$), against the direction of the flow ($b = -1$), or is not part of the cycle ($b = 0$). Tracking this derivative as described in Eq.~(8) in the main manuscript, we obtain the condition
\begin{align}
    \frac{\mathrm{d}}{\mathrm{d} u_e} \left[ \left. \frac{\partial U}{\partial x_o} \right|_{\vec{x}_o^*} \right] = b_{e o} \, l_e - \sum_{e^{\prime}} \frac{l_{e^{\prime}} \, b_{e^{\prime} o}}{1 + x_{e^{\prime}}^* + \sum_{o^{\prime\prime}} b_{e^{\prime} o^{\prime\prime}} \, x^*_{o^{\prime\prime}}} \sum_{o^{\prime}} b_{e^{\prime} o^{\prime}} \frac{\partial x_{o^{\prime}}}{\partial u_e} = 0 \,.
\end{align}
At the optimal flow state of the initial system, all cycle flows are zero, $\vec{x}^*_O=\vec{0}$, and the condition simplifies to the linear system
\begin{align}
    \sum_{o^{\prime}} \left[\sum_{e^{\prime}} \frac{l_{e^{\prime}} b_{e^{\prime} o} b_{e^{\prime} o^{\prime}}}{1 + x_{e^{\prime}}^*}\right] \frac{\partial x_{o^{\prime}}}{\partial u_e} = b_{e o} l_e \,.
\end{align}
Writing this equation in matrix notation for all links and cycles, we find
\begin{align}
    \underbrace{
        \begin{pmatrix}
            \ddots & \vdots & \iddots \\
            \cdots & \sum_{e^{\prime}} \frac{l_{e^{\prime}} b_{e^{\prime} o} b_{e^{\prime} o^{\prime}}}{1 + x_{e^{\prime}}^*} & \cdots \\
            \iddots & \vdots & \ddots \\
        \end{pmatrix}
    }_{\mathbb{R}^{(o \times o^{\prime})}}
    \cdot
    \underbrace{
        \begin{pmatrix}
            \ddots & \vdots & \iddots \\
            \cdots & \frac{\partial x_{o^{\prime}}}{\partial u_{e}} & \cdots \\
            \iddots & \vdots & \ddots \\
        \end{pmatrix}
    }_{\mathbb{R}^{(o^{\prime} \times e)}}
    =
    \underbrace{
        \begin{pmatrix}
            \ddots & \vdots & \iddots \\
            \cdots & b_{e o} l_e & \cdots \\
            \iddots & \vdots & \ddots \\
        \end{pmatrix}
    }_{\mathbb{R}^{(o \times e)}} \,.
    \label{eq:first_derivative_purc}
\end{align}
and obtain our system of linear equations for the derivatives of the cycle flows $\frac{\partial x_{o^{\prime}}}{\partial u_{e}}$. We then transform the resulting derivatives in cycle-space back into the link-space derivatives $\frac{\partial x_{e^{\prime}}^*}{\partial u_e}$ by adding the contributions from all cycles that include each link encoded by $b_{eo}$.

\newpage

\subsection*{Second-order expansion}
Applying the general second-order derivatives regarding the same link to the perturbed utility function, capturing the self-synergies of the link, we get
\begin{align}
\frac{\mathrm{d}^2 U(G,\vec{x}^*)}{\mathrm{d} u_e^2}  &= 2\,l_{e} \frac{\partial x_{e}^*}{\partial u_e} - \sum_{e^\prime\in E} l_{e^\prime} \frac{1}{1 + x_{e^\prime}^*}\left(\frac{\partial x_{e^\prime}^*}{\partial u_e}\right)^2 + \sum_{e^\prime\in E} l_{e^\prime} \left(u_{e^\prime} - \ln(1 + x_{e^\prime}^*)  \right) \frac{\partial^2 x_{e^\prime}^*}{\partial u_e^2} \,.
\end{align}
Similarly, for the second-order terms regarding different links, capturing the cross-synergies, we get
\begin{align}
    \frac{\mathrm{d}^2 U(G,\vec{x}^*)}{\mathrm{d} u_{e_2}\,\mathrm{d} u_{e_1}} &= l_{e_1} \frac{\partial x_{e_1}^*}{\partial u_{e_2}} + l_{e_2} \frac{\partial x_{e_2}^*}{\partial u_{e_1}} - \sum_{e^\prime \in E} l_{e^\prime} \frac{1}{1 + x_{e^\prime}^*}\frac{\partial x_{e^\prime}^*}{\partial u_{e_2}} \frac{\partial x_{e^\prime}^*}{\partial u_{e_1}}  \\
    &\phantom{=}\quad\quad\quad\quad\quad  + \sum_{e^\prime \in E} l_{e^\prime} \left(u_{e^\prime} - \ln(1 + x_{e^\prime}^*)  \right) \frac{\partial^2 x_{e^\prime}^*}{\partial u_{e_2} \partial u_{e_1}} \,.
\end{align}

Applying the general form from Supplementary Note 4 to the perturbed utility model results in
\begin{align}
    \frac{\mathrm{d}}{\mathrm{d} u_{e_2}}\frac{\mathrm{d}}{\mathrm{d} u_{e_1}} \left[ \left. \frac{\partial U}{\partial x_o} \right|_{\vec{x}^*} \right] &= -\frac{\mathrm{d}}{\mathrm{d} u_{e_2}} \sum_{e^\prime o^\prime} \frac{l_{e^\prime} b_{e^\prime o} b_{e^\prime o^\prime}}{1 + x_{e^\prime}^* + \sum_{o^{\prime\prime\prime}} b_{e^\prime o^{\prime\prime\prime}} x^*_{o^{\prime\prime\prime}}} \, \frac{\partial x_{o^\prime}}{\partial u_{e_1}} \\
    &= \sum_{e^\prime o^\prime} \frac{l_{e^\prime} b_{e^\prime o} b_{e^\prime o^\prime}}{(1 + x_{e^\prime}^* + \sum_{o^{\prime\prime\prime}} b_{e^\prime o^{\prime\prime\prime}} x^*_{o^{\prime\prime\prime}})^2} \, \frac{\partial x_{o^\prime}}{\partial u_{e_1}} \sum_{o^{\prime\prime}} b_{e^\prime o^{\prime\prime}} \frac{\partial x_{o^{\prime\prime}}}{\partial u_{e_2}} \\
    &\phantom{=} - \sum_{e^\prime o^\prime}\frac{l_{e^\prime} b_{e^\prime o} b_{e^\prime o^\prime}}{1 + x_{e^\prime}^* + \sum_{o^{\prime\prime\prime}} b_{e^\prime o^{\prime\prime\prime}} x^*_{o^{\prime\prime\prime}}} \frac{\partial^2 x_{o^\prime}}{\partial u_{e_1} \, \partial u_{e_2}}  \nonumber \\
    &= \sum_{e^\prime o^\prime o^{\prime\prime}} \frac{l_{e^\prime} b_{e^\prime o} b_{e^\prime o^\prime} b_{e^\prime o^{\prime\prime}}}{(1 + x_{e^\prime}^* + \sum_{o^{\prime\prime\prime}} b_{e^\prime o^{\prime\prime\prime}} x^*_{o^{\prime\prime\prime}})^2} \, \frac{\partial x_{o^\prime}}{\partial u_{e_1}} \, \frac{\partial x_{o^{\prime\prime}}}{\partial u_{e_2}} \\
    &\phantom{=} -\sum_{e^\prime o^\prime}\frac{l_{e^\prime} b_{e^\prime o} b_{e^\prime o^\prime}}{1 + x_{e^\prime}^* + \sum_{o^{\prime\prime\prime}} b_{e^\prime o^{\prime\prime\prime}} x^*_{o^{\prime\prime\prime}}} \frac{\partial^2 x_{o^\prime}}{\partial u_{e_1} \, \partial u_{e_2}} \nonumber \\
    &= 0 \,.
\end{align}
Evaluating the cycle flows at the optimal flow solution, $\vec{x}_O = \vec{0}$ as for the first derivatives, we find a linear system of equations that defines the second derivatives, given that the first derivatives are known,
\begin{equation}
    \sum_{e^\prime o^\prime o^{\prime\prime}} \frac{l_{e^\prime} b_{e^\prime o} b_{e^\prime o^\prime} b_{e^\prime o^{\prime\prime}}}{(1 + x_{e^\prime}^*)^2} \, \frac{\partial x_{o^\prime}}{\partial u_{e_2}} \, \frac{\partial x_{o^{\prime\prime}}}{\partial u_{e_1}}
    - \sum_{o^\prime e^\prime} \frac{l_{e^\prime} b_{e^\prime o} b_{e^\prime o^\prime}}{1 + x_{e^\prime}^*} \, \frac{\partial^2 x_{o^\prime}}{\partial u_{e_2} \, \partial u_{e_1}} = 0 \,.
\end{equation}

In matrix form we again find a similar set of linear equations
\begin{equation}
    \underbrace{
        \begin{pmatrix}
            \ddots & \vdots & \iddots \\
            \cdots & \sum_{e^\prime} \frac{l_{e^\prime} b_{e^\prime o} b_{e^\prime o^\prime}}{1 + x_{e^\prime}^*} & \cdots \\
            \iddots & \vdots & \ddots \\
        \end{pmatrix}
    }_{\mathbb{R}^{(o \times o^\prime)}} \cdot
    \underbrace{
        \begin{pmatrix}
            \ddots & \vdots & \iddots \\
            \cdots & \frac{\partial^2 x_{o^\prime}}{\partial u_{e_2} \, \partial u_{e_1}} & \cdots \\
            \iddots & \vdots & \ddots \\
        \end{pmatrix}
    }_{\mathbb{R}^{(o^\prime \times (e_1 \cdot e_2))}}
    =
    \underbrace{
        \begin{pmatrix}
            \ddots & \vdots & \iddots \\
            \cdots & \sum_{e^\prime o^\prime o^{\prime\prime}} \frac{l_{e^\prime} b_{e^\prime o} b_{e^\prime o^\prime} b_{e^\prime o^{\prime\prime}}}{(1 + x_{e^\prime}^*)^2} \frac{\partial x_{o^\prime}}{\partial u_{e_2}} \frac{\partial x_{o^{\prime\prime}}}{\partial u_{e_1}} & \cdots \\
            \iddots & \vdots & \ddots \\
        \end{pmatrix}
    }_{\mathbb{R}^{(o \times (e_1 \cdot e_2))}} \,,
\end{equation}
as for the first derivatives. The resulting second derivatives in cycle-space relate to the link-space derivatives in the same way as the first derivatives.

\newpage

Combining the first and second order derivatives for the  perturbed utility, we find the estimated change in utility $\Delta U(G',\vec{x}')$ for a single trip with origin $o$ and destination $d$ due to a change in utilities $u_e$ of a specific set of links $e \in \Delta E$ by some amount $\Delta u_e$
\begin{align}
    \Delta U(G',\vec{x}') &\approx \sum_{e \in \Delta E} \frac{\mathrm{d} U(G,\vec{x}^*)}{\mathrm{d} u_e} \Delta u_e + \frac{1}{2}\sum_{e \in \Delta E} \frac{\mathrm{d}^2 U(G,\vec{x}^*)}{\mathrm{d} u_e^2} \Delta u_e^2  + \frac{1}{2} \sum_{e_1 \neq e_2} \frac{\mathrm{d}^2 U(G,\vec{x}^*)}{\mathrm{d} u_{e_1}\,\mathrm{d} u_{e_2}} \Delta u_{e_1} \Delta u_{e_2} \\
    &= \sum_{e \in \Delta E}\left(l_e x_{e}^* + \sum_{n \in E} l_{n} \left( u_{n} - \ln(1 + x_{n}^*) \right) \frac{\partial x_{n}^*}{\partial u_e} \right) \Delta u_e \nonumber\\
    &\phantom{=} + \frac{1}{2}\sum_{e \in \Delta E}\left(2\,l_{e} \frac{\partial x_{e}^*}{\partial u_e} - \sum_{n\in E} l_{n} \frac{1}{1 + x_{n}^*}\left(\frac{\partial x_{n}^*}{\partial u_e}\right)^2 \right. \nonumber\\
    &\phantom{=}\quad\quad\quad\quad\quad + \left. \sum_{n\in E} l_{n} \left(u_{n} - \ln(1 + x_{n}^*)  \right) \frac{\partial^2 x_{n}^*}{\partial u_e^2} \right) \Delta u_e^2 \\
    &\phantom{=} + \frac{1}{2}\sum_{e_1 \neq e_2}\left(l_{e_1} \frac{\partial x_{e_1}^*}{\partial u_{e_2}} + l_{e_2} \frac{\partial x_{e_2}^*}{\partial u_{e_1}} - \sum_{n \in E} l_{n} \frac{1}{1 + x_{n}^*}\frac{\partial x_{n}^*}{\partial u_{e_2}} \frac{\partial x_{n}^*}{\partial u_{e_1}} \right. \nonumber \\
    &\phantom{=}\quad\quad\quad\quad\quad  + \left. \sum_{e} l_{n} \left(u_{n} - \ln(1 + x_{n}^*)  \right) \frac{\partial^2 x_{n}^*}{\partial u_{e_2} \partial u_{e_1}} \right) \Delta u_{e_1} \Delta u_{e_2} \nonumber  \,.
\end{align}

\clearpage

\section*{Supplementary Note 6: Sample calculation \\ for the perturbed utility route choice model}
\begin{figure}[!h]
    \includegraphics[scale=0.8]{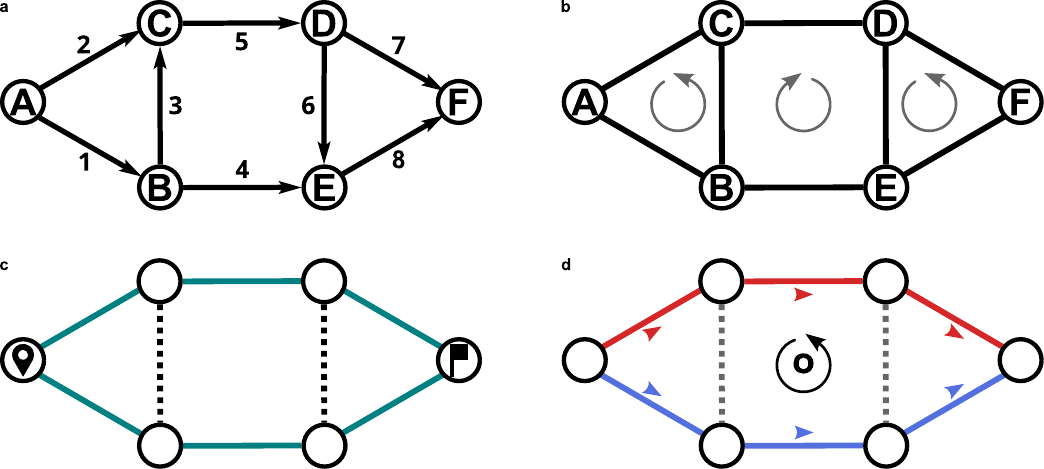}
    \caption{
    \textbf{Cycle flows explain flow changes.}
    (a) Directed network with six nodes (A to F) and eight links ($1$ to $8$).
    (b) The network consists of three basic cycles (gray arrows). All other cycles can be constructed from these three.
    (c) Unit flow from node A (pin) to F (flag) distributes equally across the upper and lower links (solid teal). The vertical links ($3$ and $6$, compare panel a) running perpendicular to the trip direction remain unused (dashed black).
    (d) The set of active links forms a single counterclockwise cycle $o$ (direction chosen arbitrarily) that describes all possible flow changes. The directed links on the bottom path are aligned with the direction of the cycle (blue), the links on the top path are anti-aligned, represented by the signed matrix $B$ [Eq.~\eqref{eq:example_cycle}].
    \vspace{-20mm}
    }
    \label{fig:SFIG2_example_network_purc}
\end{figure}

We consider a simple example network with eight directed links $e \in \{1,2,3,4,5,6,7,8\}$ and one trip from node $A$ to node $F$ with total flow $x = 1$ (Fig.~\ref{fig:SFIG2_example_network_purc}a). For simplicity, all links have a length $l_e = 1$ and utility $u_e = -1$. All flows in the network without any sources and sinks can be described as a combination of three cycle flows (Fig.~\ref{fig:SFIG2_example_network_purc}b).

Using the perturbed utility route choice model, the optimal flow for our single origin-destination pair is
\begin{align}
    X^* = \left(x_{1}, x_{2}, x_{3}  ,  x_{3}, x_{4}, x_{5}, x_{6}, x_{7}, x_{8} \right)^\mathrm{T} = \left( 1/2, 1/2, 0, 1/2, 1/2, 0, 1/2, 1/2 \right)^\mathrm{T} \,,
\end{align}
such that two links in the network remain unused (Fig.~\ref{fig:SFIG2_example_network_purc}c). Any change to this flow that does not change the set of active links is single cycle flow along the used links (Fig.~\ref{fig:SFIG2_example_network_purc}d). This cycle is described by the matrix $B$ (picking an arbitrary, but consistent direction) as
\begin{align}
    B = \begin{pmatrix}
        1\\
        -1\\
        0\\
        1\\
        -1\\
        0\\
        -1\\
        1
    \end{pmatrix} \,. \label{eq:example_cycle}
\end{align}

We obtain the first-order derivatives as computed above [Eq.~\ref{eq:first_derivative_purc}], where the matrix of the linear equation simplifies to a scalar since we have only a single cycle.
\begin{align}
    \left.\sum_{e^{\prime}} \frac{l_{e^{\prime}} b_{e^{\prime} o} b_{e^{\prime} o^{\prime}}}{1 + x_{e^{\prime}}^*}\right|_{o^{\prime}=o} = \sum_{e^{\prime}} \frac{\left(b_{e^{\prime} o}\right)^2}{1+x_{e^{\prime}}^*} = 4 \,.
\end{align}
This allows us to directly write the derivatives $\frac{\partial x_{o}}{\partial u_{e}}$ as
\begin{align}
    \left( \frac{\partial x_{o}}{\partial u_{e_{1}}}, \frac{\partial x_{o}}{\partial u_{e_{2}}}, \frac{\partial x_{o}}{\partial u_{e_{3}}}, \frac{\partial x_{o}}{\partial u_{e_{4}}}, \frac{\partial x_{o}}{\partial u_{e_{5}}}, \frac{\partial x_{o}}{\partial u_{e_{6}}},  \frac{\partial x_{o}}{\partial u_{e_{7}}}, \frac{\partial x_{o}}{\partial u_{e_{8}}} \right)
    = \frac{1}{4} \left( 1, -1, 0, 1, -1, 0, -1, 1 \right) \,,
    \label{eq:first_derivative_matrix_purc}
\end{align}
reflecting that increasing the quality of a link along the bottom path increases the cycle flow, moving flow from the top path to the upgraded bottom path (and vice versa for upgrades along the top path).

\clearpage

\section*{Supplementary Note 7: Spatial distribution of cross synergies}

\begin{figure}[!h]
    \includegraphics[width=\linewidth]{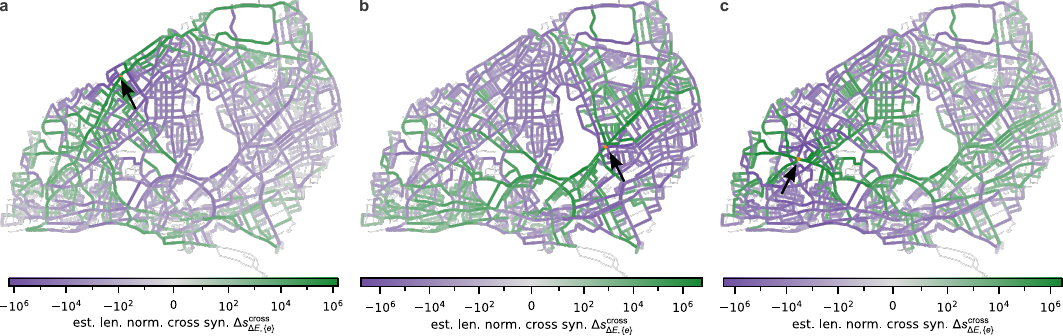}
    \caption{
    \textbf{Spatial distribution of cross-synergies in Hamburg, Germany.}
    (a-c) Estimated cross-synergy for different focal links (orange). In all examples, highly synergistic links form a funnel towards the focal link, whereas antisynergistic links provide parallel routes. The absolute value of the synergy decreases with increasing distance from the focal link (see also Supplementary Note 8).
    }
    \label{fig:SFIG3_cross_syn}
\end{figure}

\section*{Supplementary Note 8: Distance scaling of cross-synergies}

\begin{figure}[!h]
    \centering
    \includegraphics[scale=0.9]{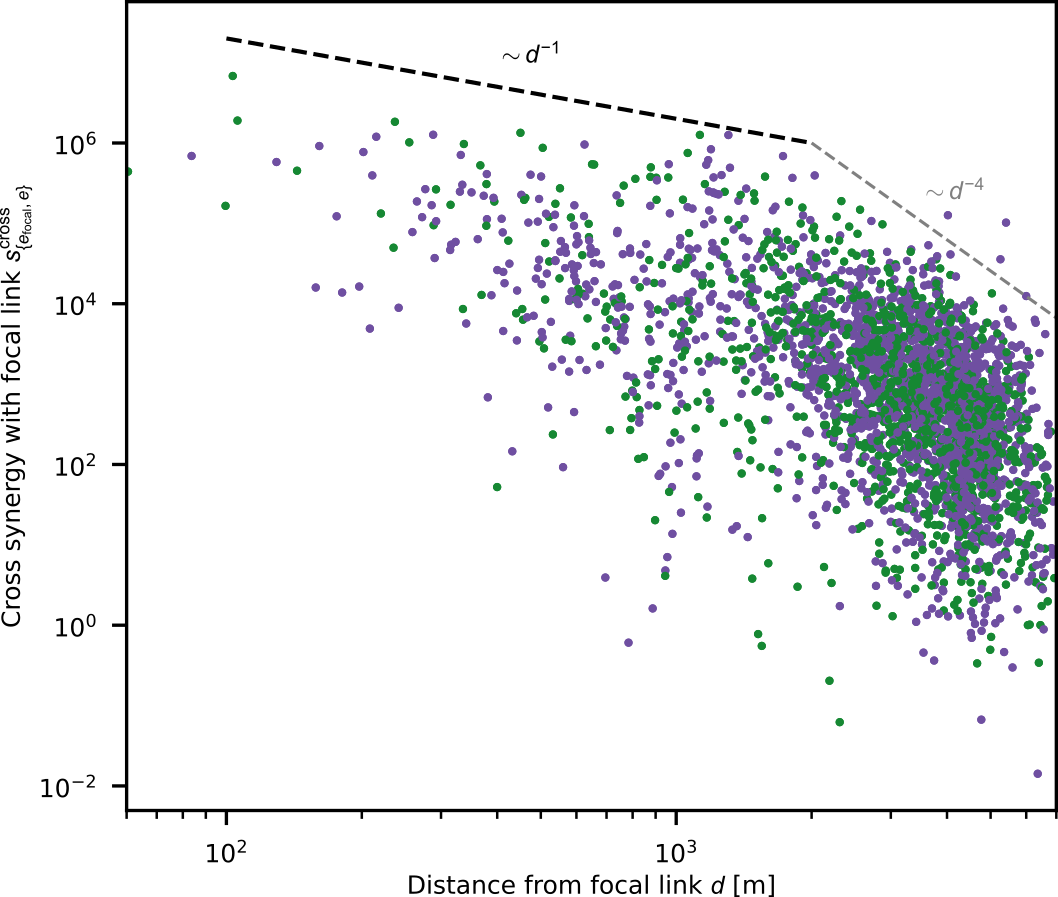}
    \caption{
    \textbf{Absolute cross-synergy decreases with distance.}
    Cross-synergies as a function of the distance between the focal link and all other links in the network for the setting shown in Fig.~3b in the main manuscript. The absolute value of the cross-synergies decreases with the distance from the focal link for both links with positive (green) and negative synergy (purple). The maximal cross-synergies seem to approximately follow simple power-law scaling. The black dashed line indicates a decay of the cross synergy inversely proportional to the distance to the focal link, $s \sim d^{-1}$ (see text). For larger distances, the scaling seems to change, and the gray dashed line indicates a decay $ \sim d^{-4}$ (no theoretical justification).
    \vspace{-20mm}
    }
    \label{fig:SFIG4_synergy_distance_scaling}
\end{figure}

The exact value of the cross-synergy between two links depends both on the distance between them and their relative orientation (e.g. whether one link routes flow towards or around the other link). However, the maximal value of the length-normalized cross-synergy seems to decay as a power law $s \sim d^{-1}$ for short distances (Fig.~\ref{fig:SFIG4_synergy_distance_scaling}). A simple geometric explanation may be the following: the value of the cross-synergy depends on how much flow the two links share. Considering a circle around one link and assuming all flow across it also crosses this circle (valid for short distances), the flow will distribute over the circumference of the circle, thus falling of inversely proportional to the distance. Consequently, the cross-synergy should also decrease inversely proportional to the distance (if the links crossing the circle are spaced equidistantly).

\clearpage

\section*{Supplementary Note 9: Network expansion (Hamburg)}

\begin{figure}[!h]
    \centering
    \includegraphics[width=\linewidth]{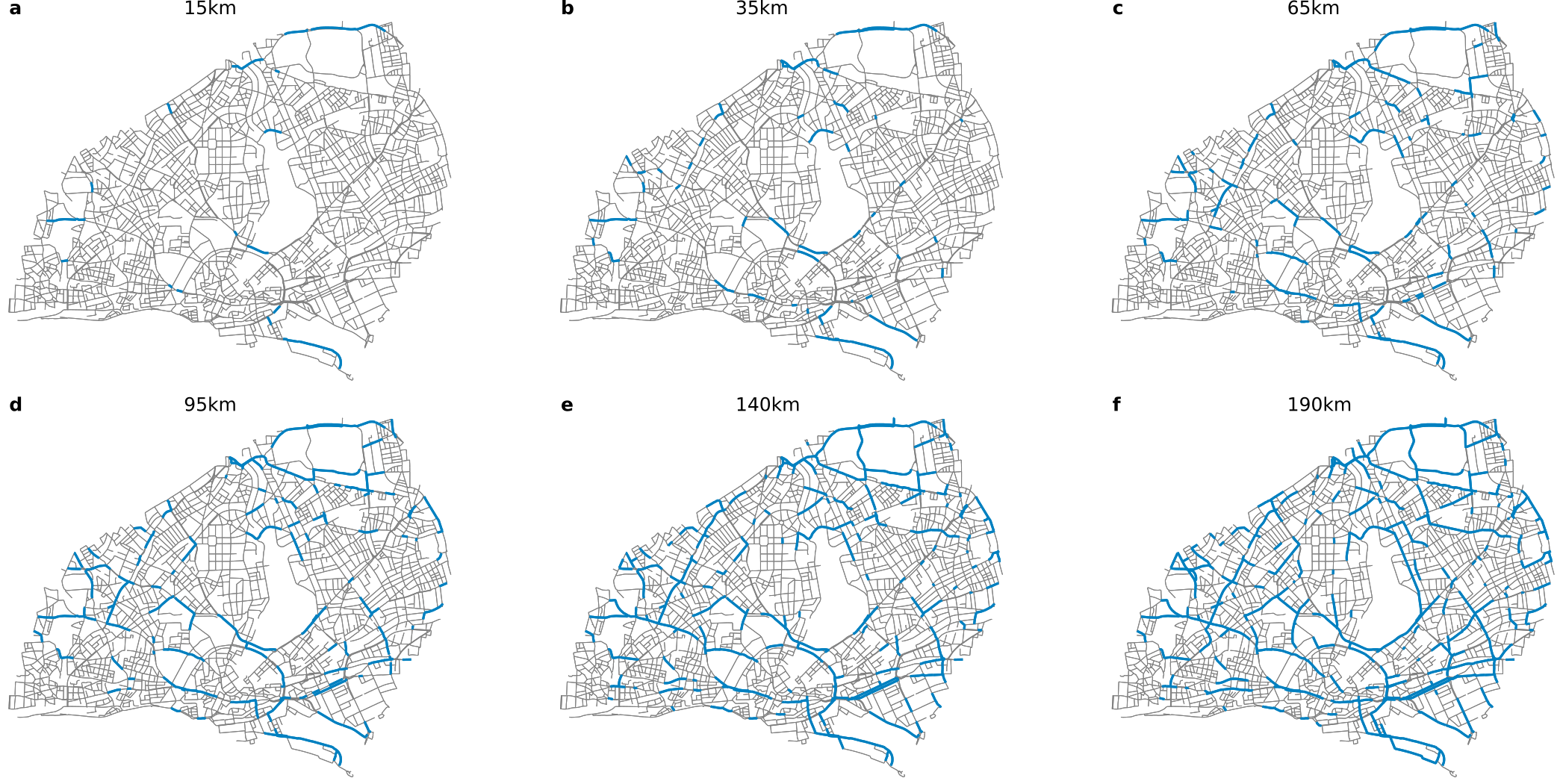}
    \caption{
    \textbf{Network expansion based on length-weighted first-order contribution.}
    Upgrading links (blue) based on their length-weighted first-order utility improvement generates a largely disconnected network but still achieves a comparatively high utility improvement due to the larger contributions of the first-order terms (compare Tab.~\ref{tab:STAB2_Hamburg_utility_improvements} and Fig.~4d in the main manuscript).
    }
    \label{fig:SFIG5_fo_util_backbone}
\end{figure}

\begin{figure}[!h]
    \centering
    \includegraphics[width=\linewidth]{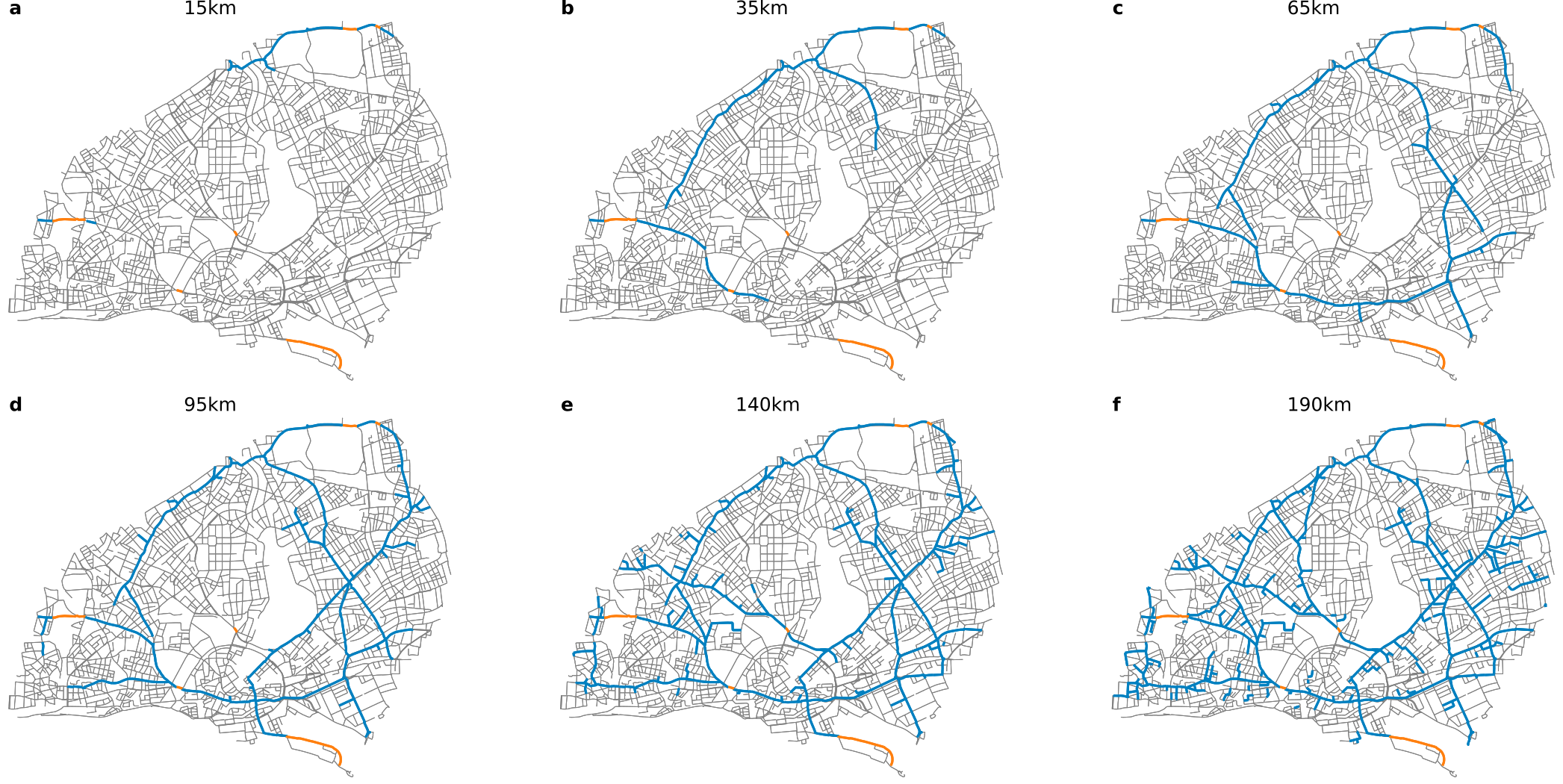}
    \caption{
    \textbf{Network expansion based on length-weighted cross-synergy contribution.}
    Starting with the ten bi-directional links (orange) with the highest first-order utility improvement and iteratively upgrading links to the bike path network (blue) based on their cross-synergy with all previous upgrades leads to a highly connected backbone, already in the early stages of the network expansion (panel c, (compare Tab.~\ref{tab:STAB2_Hamburg_utility_improvements}).
    \vspace{-20mm}
    }
    \label{fig:SFIG6_syn_backbone}
\end{figure}

\clearpage

\begin{figure}[!h]
    \centering
    \includegraphics[width=\linewidth]{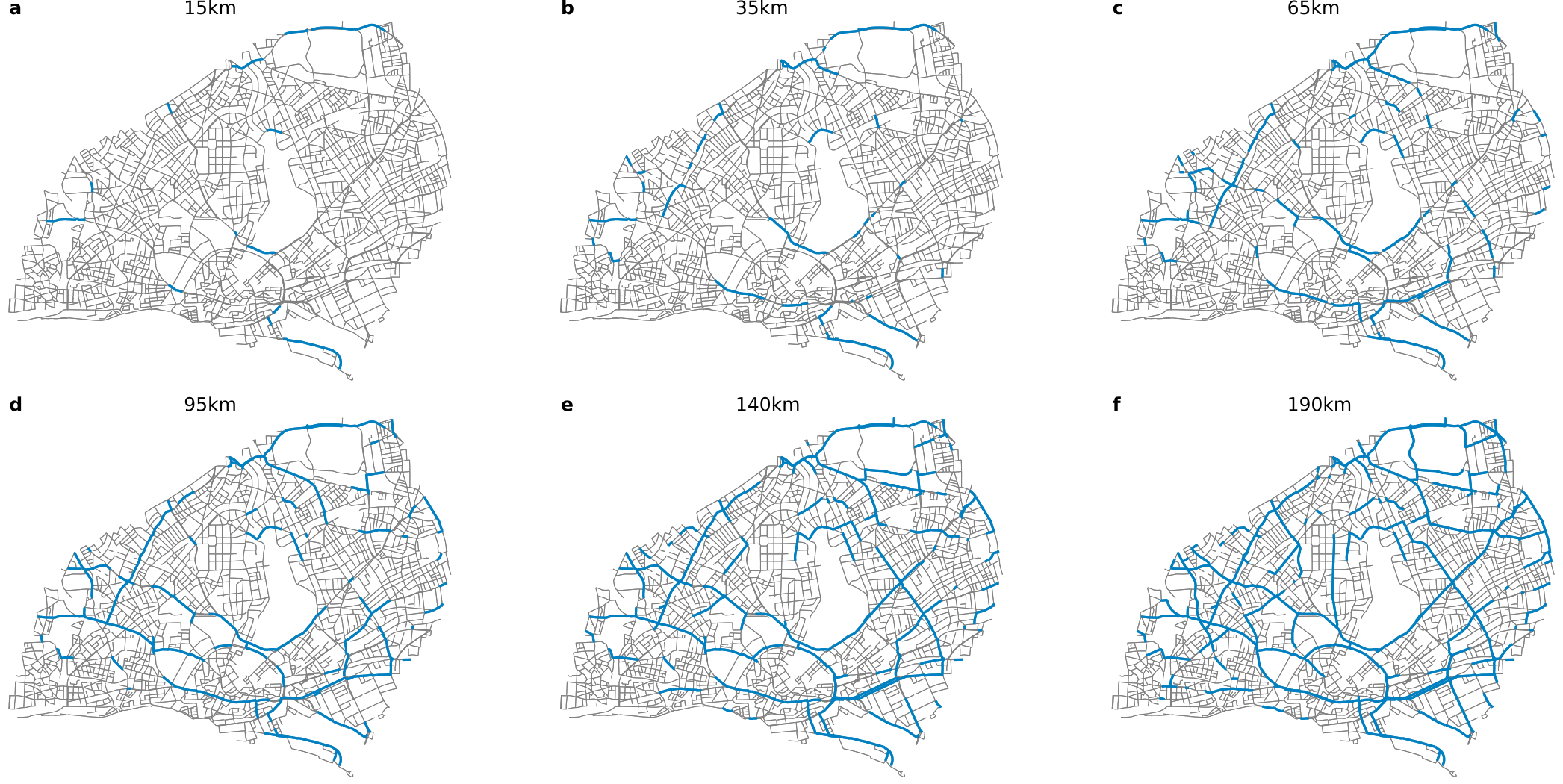}
    \caption{
    \textbf{Network expansion based on length-weighted  first- and second-order contributions.}
    While upgrading links to the bike path network (blue) based on their first- and second-order contribution results in a disconnected network in the early stages of network expansion (panels a-c), the upgrades become connected when sufficiently many links are upgraded (panel f, (compare Tab.~\ref{tab:STAB2_Hamburg_utility_improvements}).
    }
    \label{fig:SFIG7_util_backbone}
\end{figure}

\begin{table}[!h]
    \centering
    \begin{varwidth}{\columnwidth}
    \begin{tabular}{|r||c|c|c||c|c|c||c|c|c||}
         \hline
         & \multicolumn{9}{c||}{length of bike path network [km]} \\
         \hline
         & \multicolumn{3}{c||}{$15$}  & \multicolumn{3}{c||}{$35$} & \multicolumn{3}{c||}{$65$} \\
         \hline
         & fo & syn & fo+so & fo  & syn & fo+so & fo  & syn. & fo+so \\
         \hline
         Actual utility imp. & \num{0.167} & \num{0.155} & \num{0.168} & \num{0.310} & \num{0.274} & \num{0.314} & \num{0.482} & \num{0.413} & \num{0.498} \\
         Predicted utility imp. & \num{0.167} & \num{0.154} & \num{0.168} & \num{0.309} & \num{0.266} & \num{0.312} & \num{0.481} & \num{0.396} & \num{0.492} \\
         \hline
         First-order contrib. & \num{0.153} & \num{0.137} & \num{0.152} & \num{0.287} & \num{0.216} & \num{0.277} & \num{0.443} & \num{0.308} & \num{0.430} \\
         relative [\si{\percent}] & 92 & 89 & 91 & 93 & 81 & 89 & 92 & 78 & 87 \\
         \hline
         Second-order contrib. & \num{0.014} & \num{0.017} & \num{0.015} & \num{0.022} & \num{0.050} & \num{0.036} & \num{0.038} & \num{0.088} & \num{0.063} \\
         Self-synergies & \num{0.001} & \num{0.007} & \num{0.009} & \num{0.018} & \num{0.017} & \num{0.020} & \num{0.030} & \num{0.030} & \num{0.037} \\
         relative [\si{\percent}] & 5 & 5 & 5 & 6 & 6 & 7 & 6 & 7 & 7 \\
         Cross-synergies & \num{0.005} & \num{0.010} & \num{0.006} & \num{0.004} & \num{0.033} & \num{0.016} & \num{0.008} & \num{0.059} & \num{0.026} \\
         relative [\si{\percent}] & 3 & 6 & 4 & 1 & 13 & 5 & 2 & 15 & 5 \\
         \hline
         Rel. len. LCC & \num{0.22} & \num{0.61} & \num{0.22} & \num{0.17} & \num{0.65} & \num{0.14} & \num{0.11} & \num{0.95} & \num{0.20} \\
        \hline
    \end{tabular} \\[2mm]

    \begin{tabular}{|r||c|c|c||c|c|c||c|c|c||}
         \hline
         & \multicolumn{9}{c||}{length of bike path network [km]} \\
         \hline
         & \multicolumn{3}{c||}{$95$} & \multicolumn{3}{c||}{$140$} & \multicolumn{3}{c||}{$190$}\\
         \hline
         & fo & syn & fo+so & fo  & syn & fo+so & fo  & syn. & fo+so \\
         \hline
         Actual utility imp. & \num{0.621} & \num{0.523} & \num{0.645} & \num{0.800} & \num{0.640} & \num{0.827} & \num{0.972} & \num{0.707} & \num{1.000}\\
         Predicted utility imp. & \num{0.619} & \num{0.504} & \num{0.636} & \num{0.795} & \num{0.619} & \num{0.817} & \num{0.967} & \num{0.685} & \num{0.988}\\
         \hline
         First-order contrib. & \num{0.572} & \num{0.390} & \num{0.552} & \num{0.737} & \num{0.482} & \num{0.718} & \num{0.892} & \num{0.536} & \num{0.867}\\
         relative [\si{\percent}] & 92 & 77 & 87 & 93 & 78 & 88 & 92 & 78 & 88\\
         \hline
         Second-order contrib. & \num{0.047} & \num{0.114} & \num{0.084} & \num{0.058} & \num{0.137} & \num{0.099} & \num{0.074} & \num{0.149} & \num{0.120}\\
         Self-synergies & \num{0.043} & \num{0.040} & \num{0.047} & \num{0.058} & \num{0.052} & \num{0.065} & \num{0.073} & \num{0.059} & \num{0.083}\\
         relative [\si{\percent}] & 7 & 8 & 7 & 7 & 8 & 8 & 8 & 9 & 8\\
         Cross-synergies & \num{0.004} & \num{0.074} & \num{0.036} & \num{0.002} & \num{0.085} & \num{0.035} & \num{0.002} & \num{0.089} & \num{0.037}\\
         relative [\si{\percent}] & 1 & 15 & 6 & <1 & 14 & 4 & <1 & 13 & 4\\
         \hline
         Rel. len. LCC & \num{0.09} & \num{0.99} & \num{0.17} & \num{0.14} & \num{0.97} & \num{0.34} & \num{0.13} & \num{0.93} & \num{0.85}\\
        \hline
    \end{tabular}
    \end{varwidth}
    \caption{
        \textbf{Network utility improvements for the suggested network upgrades in Hamburg.}
        All values are calculated for the network states presented in Figs.~\ref{fig:SFIG5_fo_util_backbone}--\ref{fig:SFIG7_util_backbone}). Improvements of the network performance have been normalized with respect to the actual improvement with \SI{190}{\kilo\metre} network length for the first- and second-order maximizing extension model. Due to rounding effects, the individual contributions do not necessarily sum to the total predicted utility improvement.
        \vspace{-20mm}
        }
    \label{tab:STAB2_Hamburg_utility_improvements}
\end{table}

\clearpage

\section*{Supplementary Note 10: Network expansion (Manhattan)}
To illustrate the robustness of our approach, we apply the same network extension models to the street network of Manhattan. We follow the same data processing and demand modeling as outlined for Hamburg in the Methods in the main manuscript and in Supplementary Notes 1-3.

For Manhattan we keep the car-only streets outside the simulation area for the demand generation to ensure connection to Brooklyn and New Jersey via the bridges and tunnels, leaving us with a street network with $\num{2738}$ nodes and $\num{9858}$ directed links in the simulation area.

Our demand modeling area for Manhattan has $526$ cells in total, with $200$ inside the simulation area and an additional $326$ in the buffer zone. We discard $54$ cells for the same reasons as for Hamburg, leaving $\num{222000}$ unique trips in the total buffered area. $\num{37000}$ trips do not enter the simulation area at all. Of the remaining $\num{185000}$ trips, $\num{146000}$ trips are mapped into the simulation area as described for Hamburg in Supplementary Note 1. This results in approximately $\num{39000}$ unique trips within the observation area with a total of $\num{410000}$ travelers between $\num{218}$ origin/destination points.

\begin{figure}[!h]
    \centering
    \includegraphics[width=.8\linewidth]{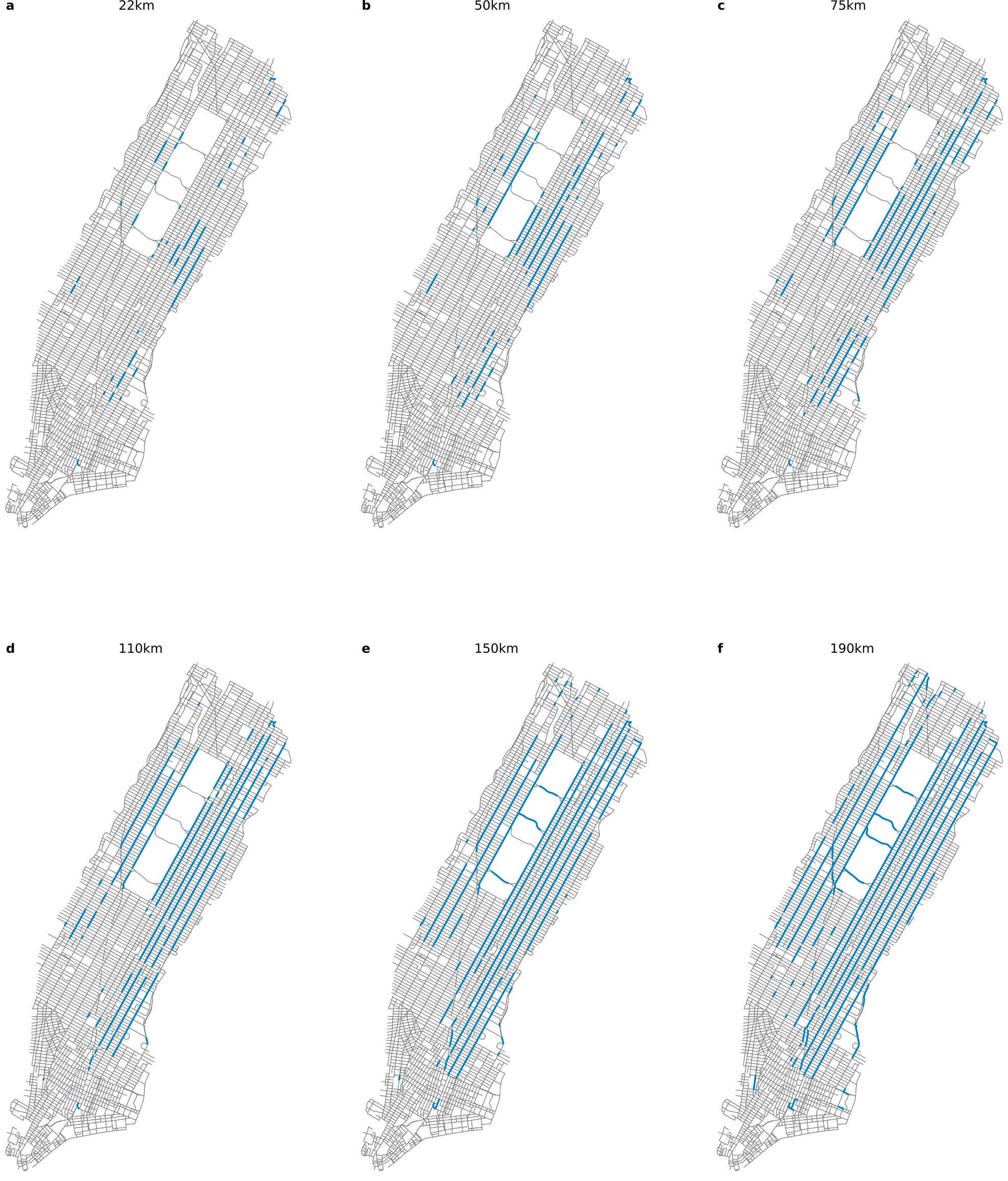}
    \caption{
    \textbf{Network expansion based on length-weighted first-order contribution.}
    Upgrading links to the bike path network (blue) based on their length-weighted first-order utility improvement generates a largely disconnected network, prioritizing links running along the long (north-south) axis of Manhattan, but still achieves a comparatively high utility improvement due to the larger contributions of the first-order terms.
    \vspace{-20mm}
    }
    \label{fig:SFIG8_ny_fo_util_backbone}
\end{figure}

\clearpage

\begin{figure}[!h]
    \centering
    \includegraphics[width=.8\linewidth]{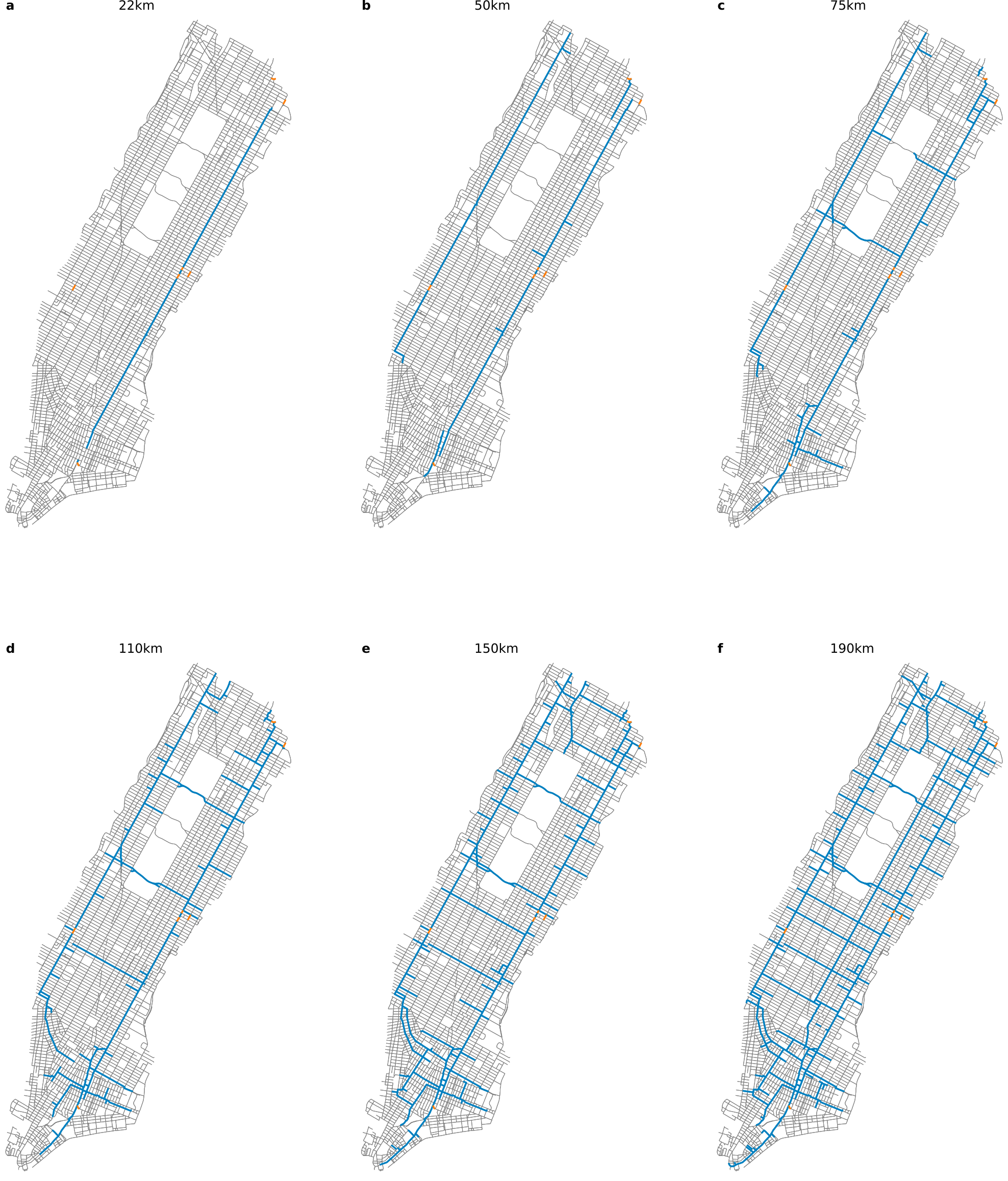}
    \caption{
    \textbf{Network expansion based on length-weighted cross-synergy contribution.}
    Starting with the ten bi-directional links (orange) with the highest first-order utility improvement and iteratively upgrading links to the bike path network (blue) based on their cross-synergy with all previous upgrades still prioritizes links running along the long (north-south) axis of Manhattan in the early stages but quickly establishes a highly connected backbone (panel c).
    }
    \label{fig:SFIG9_ny_syn_backbone}
\end{figure}

\clearpage

\begin{figure}[!h]
    \centering
    \includegraphics[width=.8\linewidth]{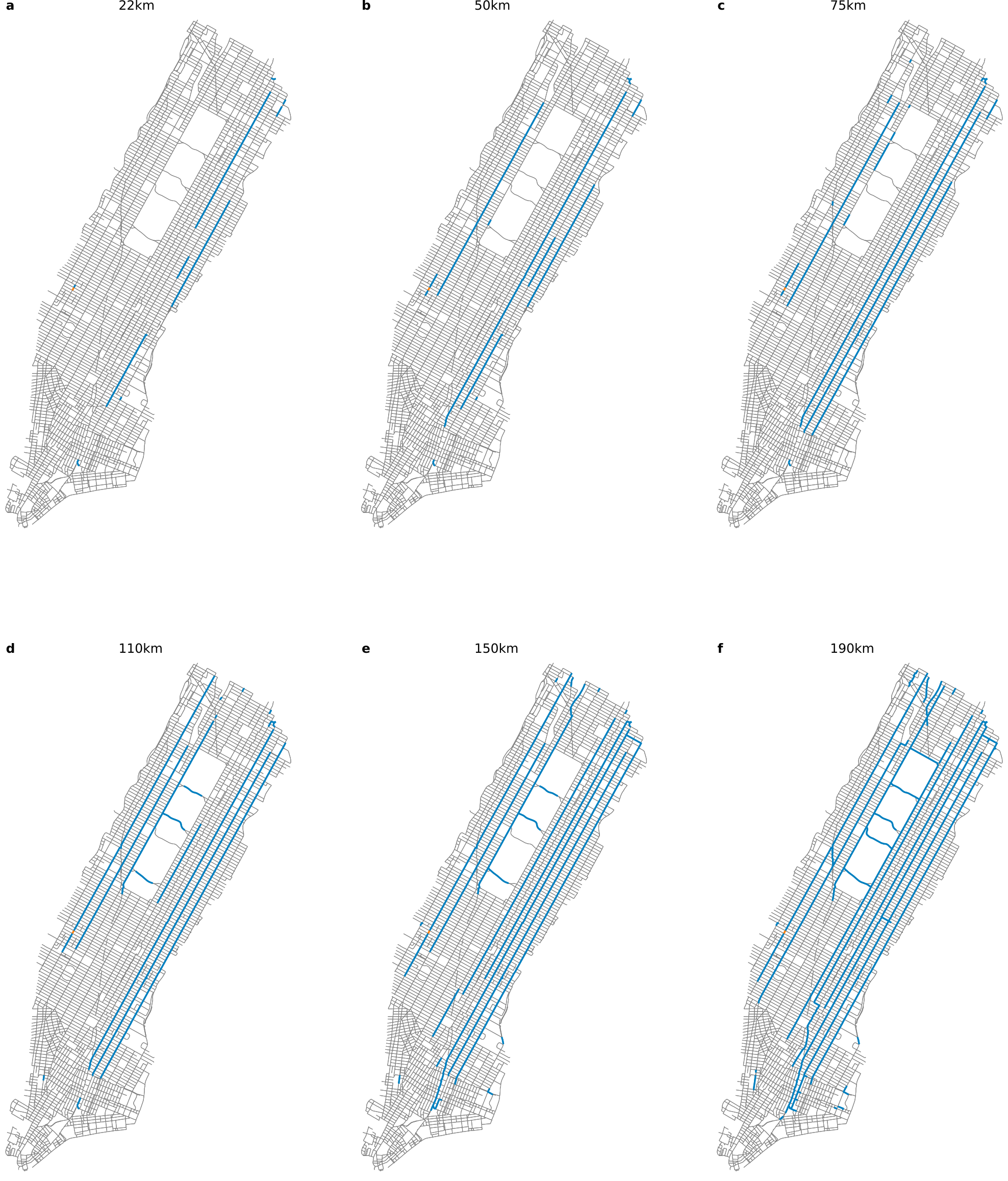}
    \caption{
    \textbf{Network expansion based on length-weighted  first- and second-order contributions.}
    While upgrading links to the bike path network (blue) based on their first- and second-order contributions still prioritizes links running along the long (north-south) axis of Manhattan but eventually achieves a slightly more connected network than first-order optimization.
   }
    \label{fig:SFIG10_ny_util_backbone}
\end{figure}

\clearpage

\section*{Supplementary Note 11: Sample calculation for traffic assignment}
In the main manuscript and Supplementary Note 5-10, we applied our approach to quantify synergies in transport infrastructure networks for non-interacting flows of cyclists computed with the perturbed utility route choice model. Here, we demonstrate the generality of the concepts by computing synergies in a small example network following a standard traffic assignment problem. We assign flows in the network determined by the Wardrop equilibrium, but evaluate the total travel time to quantify of the network performance. In contrast to the perturbed utility route choice model in our main example, we thus have two different functions $C$ and $U$ to determine network usage and quality, respectively.

\begin{figure}[!h]
    \centering
    \includegraphics[width=\linewidth]{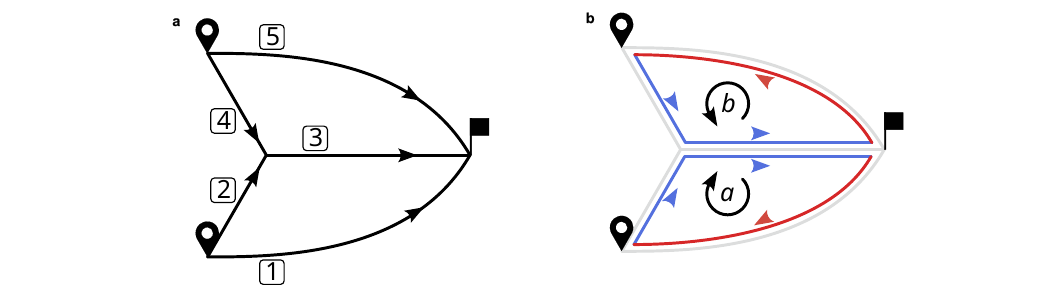}
    \caption{
    \textbf{Traffic assignment example setup.}
    (a) Simple example networks with five directed links (numbers) and two trips with different origins (pins) but the same destination (flag). Both trips have two route options in the network and only interact directly on link $3$.
    (b) Due to the limited route choices for both trips, the state of the system can be described by two cycles $a$ and $b$ with one cycle capturing the route choice decisions of each trip. We arbitrarily define the direction of the cycles such that they are aligned with the middle path (blue) and run against the flow on the outside paths (red).
    }
    \label{fig:SFIG11_sample_calculation_setup}
\end{figure}

We consider a simple network with five directed links $e \in \{1,2,3,4,5\}$ and two trips $\alpha$ and $\beta$ with different origin nodes but the same destination and total flows $x_\alpha = x_\beta = 1$ (see Fig.~\ref{fig:SFIG11_sample_calculation_setup}a). Each trip has two route options in the directed network such that the flow state of the network is described by a six-dimensional vector
\begin{equation}
    X = \left( x_{\alpha,1}, x_{\alpha,2}, x_{\alpha,3}  \;,\;  x_{\beta,3}, x_{\beta,4}, x_{\beta,5} \right)^\mathrm{T} \,,
\end{equation}
with each entry denoting the flow of the trip over one link. Drivers of the different trips only interact directly along one route on the shared link $3$.

To keep the notation consistent with the main manuscript, we formulate the problem in terms of link utilities representing the negative travel time. The utility of each link is given as
\begin{equation}
    U_e(x_e) =   \begin{cases}
                u_e\,x_e   \quad\quad e \neq 3\\
                u_e\,x_e^2 \quad\quad e = 3\\
            \end{cases} \,,
\end{equation}
with the total link flow $x_e = x_{\alpha,e} + x_{\beta,e}$ and parameters
\begin{equation}
    \left(u_e\right)_e = \left(-4, -2, -1, -2, -4\right) \,.
\end{equation}
While link $3$ has the largest utility, it is subject to congestion increasing quadratically with the total flow and mediating the interactions between the two different flows. Since we are interested in the variation of the network performance with the link utilities $u_e$, we write all expressions in the general form and only substitute the parameters when we evaluate them.

\clearpage

We measure the network performance in terms of the total utility (negative total travel time)
\begin{equation}
    U = \sum_e x_e\;U_e(x_e) = u_1\,x_{\alpha,1} + u_2\,x_{\alpha,2} + u_3\,\left(x_{\alpha,3} + x_{\beta,3}\right)^2 + u_4\,x_{\beta,4} + u_5\,x_{\beta,5} \,,
\end{equation}
and determine the equilibrium flow as the Wardrop equilibrium by maximizing
\begin{equation}
    C = \sum_e \int_0^{x_e} U_e(x^\prime)\,\mathrm{d} x^\prime = \frac{1}{2}\,u_1\,x_{\alpha,1}^2 + \frac{1}{2}\,u_2\,x_{\alpha,2}^2 + \frac{1}{3}\,u_3\,\left(x_{\alpha,3} + x_{\beta,3}\right)^3 + \frac{1}{2}\,u_4\,x_{\beta,4}^2 + \frac{1}{2}\,u_5\,x_{\beta,5}^2 \,,
\end{equation}
subject to the network flow constraints and non-negative flows. The equilibrium state in our base setting is
\begin{equation}
    X^* = \left( x_{\alpha,1}, x_{\alpha,2}, x_{\alpha,3}  ,  x_{\beta,3}, x_{\beta,4}, x_{\beta,5} \right)^\mathrm{T} = \left( 1/2, 1/2, 1/2, 1/2, 1/2, 1/2 \right)^\mathrm{T} \,,
\end{equation}
with drivers on both trips equally distributing over their respective routing options. Note that the total flow on link $3$ is still larger than on the other links, since the flow from both trips is added together, $x_3 = x_{\alpha,3} + x_{\beta,3} = 1$.

To simplify the following calculations, we express both functions $U$ and $C$ in terms of the cycle flows for the trips $\alpha$ and $\beta$ relative to the equilibrium flows $X^*$. Since each trip has two route options, we have only two cycles $a$ and $b$, with flows $x_a$ and $x_b$, reducing the number of free parameters (see Fig.~\ref{fig:SFIG11_sample_calculation_setup}b). Since we take the cycle flows around the equilibrium flows $X^*$, the equilibrium cycle flows are $x_a^* = x_b^* = 0$. We recover the absolute flows as
\begin{equation*}
    X = X^* + B\,\begin{pmatrix}x_a \\ x_b \end{pmatrix}
    \quad\quad\text{where}\quad\quad
    B = \begin{pmatrix}
            -1 & 0 \\
            1 & 0 \\
            1 & 1 \\
            0 & 1 \\
            0 & -1 \\
        \end{pmatrix} \,,
\end{equation*}
where the rows of the matrix $B$ describe the cycles $a$ and $b$, respectively, denoting which link is part of the cycle and if this link is traversed forwards ($1$) or backwards ($-1$), i.e. if increasing the cycle flow increases or decreases the flow on the link.

In terms of these cycle flows, the total network performance becomes
\begin{equation}
    U = u_1\,\left(1/2 - x_a\right) + u_2\,\left(1/2 + x_a\right) + u_3\,\left(1 + x_a + x_b\right)^2 + u_4\,\left(1/2 + x_b\right) + u_5\,\left(1/2 - x_b\right) \,,
\end{equation}
and the route choice function $C$ becomes
\begin{equation}
    C = \frac{1}{2}\,u_1\,\left(1/2 - x_a\right)^2 + \frac{1}{2}\,u_2\,\left(1/2 + x_a\right)^2 + \frac{1}{3}\,u_3\,\left(1 + x_a + x_b\right)^3 + \frac{1}{2}\,u_4\,\left(1/2 + x_b\right)^2 + \frac{1}{2}\,u_5\,\left(1/2 - x_b\right)^2 \,.
\end{equation}
We find the same equilibrium cycle flows by maximizing $C$ subject to positive link flows, such that $-1/2 \le x_a, x_b \le 1/2$. The network flow constraints are automatically fulfilled. Importantly, we can compute all derivatives and track the change of the equilibrium flows as long as the flows distribute over both route options.

We focus here on the impact of link upgrades to link $2$ and its synergies with other links. To evaluate the expected network performance changes $\Delta U$, we compute its derivatives. We find the first derivatives by tracking the equilibrium flow solution with $\frac{\partial C}{\partial x_a} = \frac{\partial C}{\partial x_b} = 0$ as discussed in the main manuscript [Eq.~(9)],
\begin{eqnarray}
    0 = \frac{\mathrm{d}}{\mathrm{d}u_2}\,\left[\frac{\partial C}{\partial x_a}\right]_{X^*} &=& \left.\frac{\partial^2 C}{\partial u_2\,\partial x_a}\right|_{X^*} + \left.\frac{\partial^2 C}{\partial x_a^2}\,\frac{\partial x_a}{\partial u_2}\right|_{X^*} + \left.\frac{\partial^2 C}{\partial x_b\,\partial x_a}\,\frac{\partial x_b}{\partial u_2}\right|_{X^*} \\
    &=& \frac{1}{2} + \Big[u_1 + u_2 + 2\,u_3\Big]\,\left.\frac{\partial x_a}{\partial u_2}\right|_{X^*} + \Big[2\,u_3\Big]\,\left.\frac{\partial x_b}{\partial u_2}\right|_{X^*} \nonumber\\[4mm]
    0 = \frac{\mathrm{d}}{\mathrm{d}u_2}\,\left[\frac{\partial C}{\partial x_b}\right]_{X^*} &=& \left.\frac{\partial^2 C}{\partial u_2\,\partial x_b}\right|_{X^*} + \left.\frac{\partial^2 C}{\partial x_a \, \partial x_b}\,\frac{\partial x_a}{\partial u_2}\right|_{X^*} + \left.\frac{\partial^2 C}{\partial x_b^2}\,\frac{\partial x_b}{\partial u_2}\right|_{X^*} \\
    &=& 0 + \Big[2\,u_3\Big]\,\left.\frac{\partial x_a}{\partial u_2}\right|_{X^*} + \Big[2\,u_3 + u_4 + u_5\Big]\,\left.\frac{\partial x_b}{\partial u_2}\right|_{X^*}  \,.\nonumber
\end{eqnarray}
The constant term in the second equation is zero since cycle $b$ does not contain link $2$ and consequently $C$ does not contain a term simultaneously depending directly on both $x_b$ and $u_e$. Writing this equation in matrix form
\begin{equation}
    \begin{pmatrix}
        u_1 + u_2 + 2\,u_3 & 2\,u_3 \\
        2\,u_3 & 2\,u_3 + u_4 + u_5
    \end{pmatrix}\,
    \begin{pmatrix}
        \frac{\partial x_a}{\partial u_2} \\ \frac{\partial x_b}{\partial u_2}
    \end{pmatrix} =
    \begin{pmatrix}
        -1/2 \\ 0
    \end{pmatrix} \,,
\end{equation}
we obtain the derivatives as the solution of this a system of two linear equations as
\begin{equation}
    \begin{pmatrix}
        \frac{\partial x_a}{\partial u_2} \\ \frac{\partial x_b}{\partial u_2}
    \end{pmatrix} =
    \begin{pmatrix}
        1/15 \\ -1/60
    \end{pmatrix} \,.
\end{equation}
As expected, improving link $2$ increases the flow along cycle $a$ and thereby the flow over link $2$. Due to the interaction between the two trips, the flow in cycle $b$ decreases slightly since the travel time along link $3$ increases.

We obtain the other derivatives with an identical calculation in a single large system of equations
\begin{equation}
    \begin{pmatrix}
        u_1 + u_2 + 2\,u_3 & 2\,u_3 \\
        2\,u_3 & 2\,u_3 + u_4 + u_5
    \end{pmatrix}\,
    \begin{pmatrix}
        \frac{\partial x_a}{\partial u_1} & \frac{\partial x_a}{\partial u_2} & \frac{\partial x_a}{\partial u_3} & \frac{\partial x_a}{\partial u_4} & \frac{\partial x_a}{\partial u_5} \\ \frac{\partial x_b}{\partial u_1} & \frac{\partial x_b}{\partial u_2} & \frac{\partial x_b}{\partial u_3} & \frac{\partial x_b}{\partial u_4} & \frac{\partial x_b}{\partial u_5}
    \end{pmatrix} =
    \begin{pmatrix}
        1/2 & -1/2 & -1 & 0 & 0 \\ 0 & 0 & -1 & -1/2 & 1/2
    \end{pmatrix} \,,
\end{equation}
as
\begin{equation}
    \begin{pmatrix}
        \frac{\partial x_a}{\partial u_1} & \frac{\partial x_a}{\partial u_2} & \frac{\partial x_a}{\partial u_3} & \frac{\partial x_a}{\partial u_4} & \frac{\partial x_a}{\partial u_5} \\ \frac{\partial x_b}{\partial u_1} & \frac{\partial x_b}{\partial u_2} & \frac{\partial x_b}{\partial u_3} & \frac{\partial x_b}{\partial u_4} & \frac{\partial x_b}{\partial u_5}
    \end{pmatrix} =
    \begin{pmatrix}
        -1/15 & 1/15 & 1/10 & -1/60 & 1/60 \\ 1/60 & -1/60 & 1/10 & 1/15 & -1/15
    \end{pmatrix} \,.
\end{equation}
The coefficients in the matrix remain the same for all derivatives and simply capture the overlap between the cycles. The differences between the derivatives are solely determined by the constant term, reflecting which cycle the upgraded link is part of.

We compute the second derivatives following the same procedure, here explicitly for $\frac{\partial^2 x_a}{\partial u_2\,\partial u_4}$ and without the reminder that the derivatives are evaluated at the equilibrium flows to simplify notation,
\begin{eqnarray}
    0 &=&  \frac{\mathrm{d}}{\mathrm{d}u_4}\left[\frac{\mathrm{d}}{\mathrm{d}u_2}\,\left[\frac{\partial C}{\partial x_a}\right]\right] \\
    &=& \phantom{+} \, \frac{\partial^3 C}{\partial u_2\,\partial u_4\,\partial x_a} + \frac{\partial^3 C}{\partial u_4\,\partial x_a^2}\,\frac{\partial x_a}{\partial u_2} + \frac{\partial^2 C}{\partial x_a^2}\,\frac{\partial^2 x_a}{\partial u_2\,\partial u_4} + \frac{\partial^3 C}{\partial u_4\,\partial x_a\,\partial x_b}\,\frac{\partial x_b}{\partial u_2} + \frac{\partial^2 C}{\partial x_a\,\partial x_b}\,\frac{\partial^2 x_b}{\partial u_2\,\partial u_4} \nonumber \\
    && + \,
    \frac{\partial^3 C}{\partial u_2\,\partial x_a^2}\,\frac{\partial x_a}{\partial u_4} + \frac{\partial^3 C}{\partial x_a^3}\,\frac{\partial x_a}{\partial u_2}\,\frac{\partial x_a}{\partial u_4} + \frac{\partial^2 C}{\partial x_a^2}\,\frac{\partial^2 x_a}{\partial u_2\,\partial x_a}\,\frac{\partial x_a}{\partial u_4} + \frac{\partial^3 C}{\partial x_a^2\,\partial x_b}\,\frac{\partial x_b}{\partial u_2}\,\frac{\partial x_a}{\partial u_4} + \frac{\partial^2 C}{\partial x_a\,\partial x_b}\,\frac{\partial^2 x_b}{\partial u_2\,\partial x_a}\,\frac{\partial x_a}{\partial u_4} \nonumber \\
    && + \,
    \frac{\partial^3 C}{\partial u_2\,\partial x_a\,\partial x_b}\,\frac{\partial x_b}{\partial u_4} + \frac{\partial^3 C}{\partial x_a^2\,\partial x_b}\,\frac{\partial x_a}{\partial u_2}\,\frac{\partial x_b}{\partial u_4} + \frac{\partial^2 C}{\partial x_a^2}\,\frac{\partial^2 x_a}{\partial u_2\,\partial x_b}\,\frac{\partial x_b}{\partial u_4} + \frac{\partial^3 C}{\partial x_a\,\partial x_b^2}\,\frac{\partial x_b}{\partial u_2}\,\,\frac{\partial x_b}{\partial u_4} + \frac{\partial^2 C}{\partial x_a\,\partial x_b}\,\frac{\partial^2 x_b}{\partial u_2\,\partial x_b}\,\frac{\partial x_b}{\partial u_4} \nonumber \\
    &=& \phantom{+} \, 0 + 0 + \left[u_1 + u_2 + 2\,u_3\right]\,\frac{\partial^2 x_a}{\partial u_2\,\partial u_4} + 0 + 2\,u_3\,\frac{\partial^2 x_b}{\partial u_2\,\partial u_4} \nonumber\\
    && +\, \frac{\partial x_a}{\partial u_4} + 2\,u_3\,\frac{\partial x_a}{\partial u_2}\,\frac{\partial x_a}{\partial u_4} + 0 + 2\,u_3\,\frac{\partial x_b}{\partial u_2}\,\frac{\partial x_a}{\partial u_4} + 0 \nonumber\\
    && +\, 0 + 2\,u_3\,\frac{\partial x_a}{\partial u_2}\,\frac{\partial x_b}{\partial u_4} + 0 + 2\,u_3\,\frac{\partial x_b}{\partial u_2}\,\frac{\partial x_b}{\partial u_4} + 0 \nonumber\\
    &=& \left[ - \frac{1}{60} - \frac{2\,u_3}{15\times 60} + \frac{ 2\,u_3}{15 \times 15} + \frac{ 2\,u_3}{60 \times 60} - \frac{2\,u_3}{ 60\times 15}\right] + \left[u_1 + u_2 + 2\,u_3\right]\,\frac{\partial^2 x_a}{\partial u_2\,\partial u_4} + \left[2\,u_3\right]\,\frac{\partial^2 x_b}{\partial u_2\,\partial u_4} \,. \nonumber
\end{eqnarray}
The very first term is zero since the route choice function $C$ is linear in the link utilities and $u_2$ and $u_4$ never appear together in the same term. Similarly, third derivatives of $C$ involving both $x_a$ and $x_b$ together with one of link utilities do not contribute since the flows only directly interact along link $3$, which we do not consider here. Finally, terms of the form $\frac{\partial^2 x_j}{\partial u_2\,\partial x_k}$ do not contribute since $\frac{\partial x_j}{\partial x_k} = \delta_{j,k}$ is constant and its derivative is zero. In the end, we again find a system of linear equations with the same matrix as for the first derivatives but different constant terms depending on the first derivatives, for example
\begin{equation}
    \begin{pmatrix}
        u_1 + u_2 + 2\,u_3 & 2\,u_3 \\
        2\,u_3 & 2\,u_3 + u_4 + u_5
    \end{pmatrix}\,
    \begin{pmatrix}
         \frac{\partial^2 x_a}{\partial u_2\,\partial u_4} \\ \frac{\partial^2 x_b}{\partial u_2\,\partial u_4}
    \end{pmatrix} =
    -\begin{pmatrix}
        - \frac{1}{60} - \frac{2\,u_3}{15\times 60} + \frac{ 2\,u_3}{15 \times 15} + \frac{ 2\,u_3}{60 \times 60} - \frac{2\,u_3}{ 60\times 15} \\
        - \frac{1}{60} - \frac{2\,u_3}{15\times 60} + \frac{ 2\,u_3}{15 \times 15} + \frac{ 2\,u_3}{60 \times 60} - \frac{2\,u_3}{ 60\times 15} \,.
    \end{pmatrix}
\end{equation}
By solving the linear equations for all second derivatives we obtain
\begin{eqnarray}
    \begin{pmatrix}
        \frac{\partial^2 x_a}{\partial u_1\,\partial u_1} & \frac{\partial^2 x_a}{\partial u_1\,\partial u_2} & \frac{\partial^2 x_a}{\partial u_1 \, \partial u_3} & \frac{\partial^2 x_a}{\partial u_1 \, \partial u_4} & \frac{\partial^2 x_a}{\partial u_1 \, \partial u_5} \\[1mm]
        \frac{\partial^2 x_a}{\partial u_2 \, \partial u_1} & \frac{\partial^2 x_a}{\partial u_2 \, \partial u_2} & \frac{\partial^2 x_a}{\partial u_2 \, \partial u_3} & \frac{\partial^2 x_a}{\partial u_2 \, \partial u_4} & \frac{\partial^2 x_a}{\partial u_2 \, \partial u_5} \\[1mm] \frac{\partial^2 x_a}{\partial u_3 \, \partial u_1} & \frac{\partial^2 x_a}{\partial u_3 \, \partial u_2} & \frac{\partial^2 x_a}{\partial u_3 \, \partial u_3} & \frac{\partial^2 x_a}{\partial u_3 \, \partial u_4} & \frac{\partial^2 x_a}{\partial u_3 \, \partial u_5} \\[1mm]
        \frac{\partial^2 x_a}{\partial u_4 \, \partial u_1} & \frac{\partial^2 x_a}{\partial u_4 \, \partial u_2} & \frac{\partial^2 x_a}{\partial u_4 \, \partial u_3} & \frac{\partial^2 x_a}{\partial u_4 \, \partial u_4} & \frac{\partial^2 x_a}{\partial u_4 \, \partial u_5} \\[1mm]
        \frac{\partial^2 x_a}{\partial u_5 \, \partial u_1} & \frac{\partial^2 x_a}{\partial u_5 \, \partial u_2} & \frac{\partial^2 x_a}{\partial u_5 \, \partial u_3} & \frac{\partial^2 x_a}{\partial u_5 \, \partial u_4} & \frac{\partial^2 x_a}{\partial u_5 \, \partial u_5}
    \end{pmatrix}
    &=&
    \begin{pmatrix}
        -\frac{329}{18000} & \phantom{-}\frac{9}{18000} & \phantom{-}\frac{96}{18000} & -\frac{41}{18000} & \phantom{-}\frac{21}{18000} \\[1mm]
        \phantom{-}\frac{9}{18000} & \phantom{-}\frac{311}{18000} & \phantom{-}\frac{384}{18000} & -\frac{39}{18000} & \phantom{-}\frac{59}{18000} \\[1mm]
        \phantom{-}\frac{96}{18000} & \phantom{-}\frac{384}{18000} & \phantom{-}\frac{1296}{18000} & \phantom{-}\frac{84}{18000} & -\frac{204}{18000} \\[1mm]
        -\frac{41}{18000} & -\frac{39}{18000} & \phantom{-}\frac{84}{18000} & -\frac{89}{18000} & \phantom{-}\frac{9}{18000} \\[1mm]
        \phantom{-}\frac{7}{6000} & \phantom{-}\frac{59}{18000} & -\frac{204}{18000} & \phantom{-}\frac{9}{18000} & \phantom{-}\frac{71}{18000}
    \end{pmatrix}
    \\
    \begin{pmatrix}
        \frac{\partial^2 x_b}{\partial u_1\,\partial u_1} & \frac{\partial^2 x_b}{\partial u_1\,\partial u_2} & \frac{\partial^2 x_b}{\partial u_1 \, \partial u_3} & \frac{\partial^2 x_b}{\partial u_1 \, \partial u_4} & \frac{\partial^2 x_b}{\partial u_1 \, \partial u_5} \\[1mm]
        \frac{\partial^2 x_b}{\partial u_2 \, \partial u_1} & \frac{\partial^2 x_b}{\partial u_2 \, \partial u_2} & \frac{\partial^2 x_b}{\partial u_2 \, \partial u_3} & \frac{\partial^2 x_b}{\partial u_2 \, \partial u_4} & \frac{\partial^2 x_b}{\partial u_2 \, \partial u_5} \\[1mm] \frac{\partial^2 x_b}{\partial u_3 \, \partial u_1} & \frac{\partial^2 x_b}{\partial u_3 \, \partial u_2} & \frac{\partial^2 x_b}{\partial u_3 \, \partial u_3} & \frac{\partial^2 x_b}{\partial u_3 \, \partial u_4} & \frac{\partial^2 x_b}{\partial u_3 \, \partial u_5} \\[1mm]
        \frac{\partial^2 x_b}{\partial u_4 \, \partial u_1} & \frac{\partial^2 x_b}{\partial u_4 \, \partial u_2} & \frac{\partial^2 x_b}{\partial u_4 \, \partial u_3} & \frac{\partial^2 x_b}{\partial u_4 \, \partial u_4} & \frac{\partial^2 x_b}{\partial u_4 \, \partial u_5} \\[1mm]
        \frac{\partial^2 x_b}{\partial u_5 \, \partial u_1} & \frac{\partial^2 x_b}{\partial u_5 \, \partial u_2} & \frac{\partial^2 x_b}{\partial u_5 \, \partial u_3} & \frac{\partial^2 x_b}{\partial u_5 \, \partial u_4} & \frac{\partial^2 x_b}{\partial u_5 \, \partial u_5}
    \end{pmatrix}
    &=&
    \begin{pmatrix}
        \phantom{-}\frac{71}{18000} & \phantom{-}\frac{9}{18000} & -\frac{204}{18000} & \phantom{-}\frac{59}{18000} & \phantom{-}\frac{21}{18000} \\[1mm]
        \phantom{-}\frac{9}{18000} & -\frac{89}{18000} & \phantom{-}\frac{84}{18000} & -\frac{39}{18000} & -\frac{41}{18000} \\[1mm]
        \frac{204}{18000} & \phantom{-}\frac{84}{18000} & \phantom{-}\frac{1296}{18000} & \phantom{-}\frac{384}{18000} & \phantom{-}\frac{96}{18000} \\[1mm]
        \phantom{-}\frac{59}{18000} & -\frac{39}{18000} & \phantom{-}\frac{384}{18000} & \phantom{-}\frac{311}{18000} & \phantom{-}\frac{9}{18000} \\[1mm]
        \phantom{-}\frac{21}{18000} & -\frac{41}{18000} & \phantom{-}\frac{96}{18000} & \phantom{-}\frac{9}{18000} & -\frac{329}{18000}
    \end{pmatrix} \,. \nonumber
\end{eqnarray}

\begin{figure}[!h]
    \centering
    \includegraphics{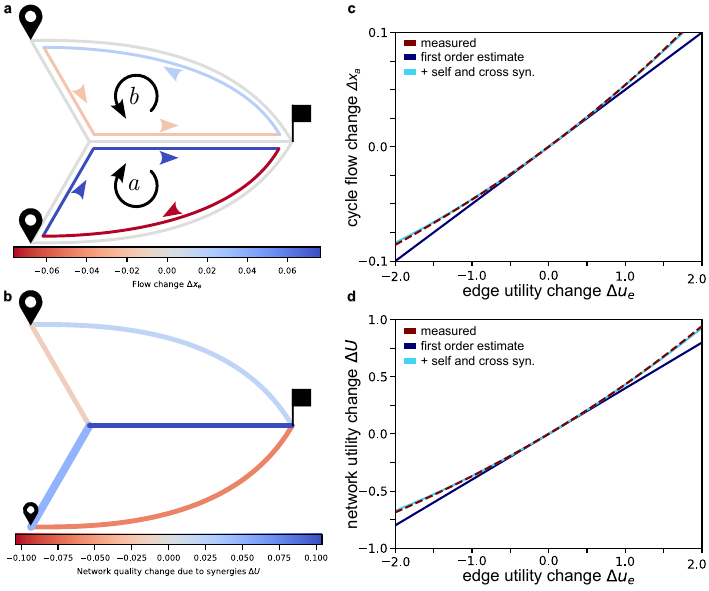}
    \caption{
    \textbf{Flow changes and synergies in traffic assignment.}
    (a) Link flow changes in the network when upgrading link $2$ (left bottom) from $u_2 = -2$ to $u_2 = -1$. Flow of trip $\alpha$ (bottom) is encouraged to route through the middle path, increasing travel times and partially displacing the flow of trip $\beta$ (top) to the outside path.
    (b) Synergies of link upgrades with link $2$ [compare row $2$ of the matrix in Eq.~\ref{eq:trafficassignment_synergies}]. Synergistic links further increase the benefits of the upgrades or reduce the negative interactions between the two flows.
    (c) Cycle flows changes $\Delta x_a$ when upgrading link $2$.
    (d) Total network performance change $\Delta U$ when upgrading link $2$. Second-order synergies are crucial to correctly capture the effects on flows and network performance.
    }
    \label{fig:SFIG12_sample_calculation_results}
\end{figure}

Substituting these derivatives, we obtain the first and second derivatives of the network performance $U$
\begin{eqnarray}
    \begin{pmatrix}
        \frac{\partial U}{\partial u_1} & \frac{\partial U}{\partial u_2} & \frac{\partial U}{\partial u_3} & \frac{\partial U}{\partial u_4} & \frac{\partial U}{\partial u_5}
    \end{pmatrix}
    &=&
    \begin{pmatrix}
        \frac{3}{10} & \frac{2}{10} & \frac{8}{10} & \frac{2}{10} & \frac{3}{10}
    \end{pmatrix}
    \label{eq:trafficassignment_importance} \\[3mm]
    \begin{pmatrix}
        \frac{\partial^2 U}{\partial u_1\,\partial u_1} & \frac{\partial^2 U}{\partial u_1\,\partial u_2} & \frac{\partial^2 U}{\partial u_1\,\partial u_3} & \frac{\partial^2 U}{\partial u_1\,\partial u_4} & \frac{\partial^2 U}{\partial u_1\,\partial u_5} \\[1mm]
        \frac{\partial^2 U}{\partial u_2\,\partial u_1} & \frac{\partial^2 U}{\partial u_2\,\partial u_2} & \frac{\partial^2 U}{\partial u_2\,\partial u_3} & \frac{\partial^2 U}{\partial u_2\,\partial u_4} & \frac{\partial^2 U}{\partial u_2\,\partial u_5} \\[1mm]
        \frac{\partial^2 U}{\partial u_3\,\partial u_1} & \frac{\partial^2 U}{\partial u_3\,\partial u_2} & \frac{\partial^2 U}{\partial u_3\,\partial u_3} & \frac{\partial^2 U}{\partial u_3\,\partial u_4} & \frac{\partial^2 U}{\partial u_3\,\partial u_5} \\[1mm]
        \frac{\partial^2 U}{\partial u_4\,\partial u_1} & \frac{\partial^2 U}{\partial u_4\,\partial u_2} & \frac{\partial^2 U}{\partial u_4\,\partial u_3} & \frac{\partial^2 U}{\partial u_4\,\partial u_4} & \frac{\partial^2 U}{\partial u_4\,\partial u_5} \\[1mm]
        \frac{\partial^2 U}{\partial u_5\,\partial u_1} & \frac{\partial^2 U}{\partial u_5\,\partial u_2} & \frac{\partial^2 U}{\partial u_5\,\partial u_3} & \frac{\partial^2 U}{\partial u_5\,\partial u_4} & \frac{\partial^2 U}{\partial u_5\,\partial u_5}
    \end{pmatrix}
    &=&
    \begin{pmatrix}
        \phantom{-}\frac{114}{1500} & -\frac{94}{1500} & -\frac{186}{1500} & \phantom{-}\frac{31}{1500} & -\frac{36}{1500} \\[1mm]
        -\frac{94}{1500} & \phantom{-}\frac{74}{1500} & \phantom{-}\frac{156}{1500} & -\frac{26}{1500} & \phantom{-}\frac{31}{1500} \\[1mm]
        -\frac{186}{1500} & \phantom{-}\frac{156}{1500} & \phantom{-}\frac{864}{1500} & \phantom{-}\frac{156}{1500} & -\frac{186}{1500} \\[1mm]
        \phantom{-}\frac{31}{1500} & -\frac{26}{1500} & \phantom{-}\frac{156}{1500} & \phantom{-}\frac{74}{1500} & -\frac{95}{1500} \\[1mm]
        -\frac{36}{1500} & \phantom{-}\frac{31}{1500} & -\frac{186}{1500} & -\frac{94}{1500} & \phantom{-}\frac{114}{1500}
    \end{pmatrix} \,. \label{eq:trafficassignment_synergies}
\end{eqnarray}
Evaluating the synergies between all links, we find the expected pattern (Fig.~\ref{fig:SFIG12_sample_calculation_results}a,b). Upgrading the main link $3$ is strongly synergistic since it affects both trip flows and reduces the negative interactions. Upgrading parallel routes for the same flow is antisynergistic for the overall network performance. However, in contrast to the non-interacting flows in our main cycling examples, upgrading alternative routes for the other origin-destination pair (e.g. upgrading both link $2$ and link $5$) is also synergistic, encouraging the two trip flows to use different routes and reducing negative interactions. As with our examples in the main manuscript, the second-order expansion accurately predicts the flow and network performance changes (Fig.~\ref{fig:SFIG12_sample_calculation_results}c,d).

\end{document}